\documentclass[12pt]{article}
\usepackage[numbers,super,square,sort&compress]{natbib}
\usepackage{graphicx}
\usepackage{fullpage}
\usepackage{amsmath}
\usepackage{threeparttable}
\usepackage{booktabs}
\usepackage{hyperref}
\usepackage{times}
\usepackage{color}           

\newcommand{\spacing}[1]{\renewcommand{\baselinestretch}{#1}\large\normalsize}
\spacing{1.5}

\newenvironment{affiliations}{%
    \setcounter{enumi}{1}%
    \setlength{\parindent}{0in}%
    \slshape\sloppy%
    \begin{list}{\upshape$^{\arabic{enumi}}$}{%
        \usecounter{enumi}%
        \setlength{\leftmargin}{0in}%
        \setlength{\topsep}{0in}%
        \setlength{\labelsep}{0in}%
        \setlength{\labelwidth}{0in}%
        \setlength{\listparindent}{0in}%
        \setlength{\itemsep}{0ex}%
        \setlength{\parsep}{0in}%
        }
    }{\end{list}\par\vspace{12pt}}

\renewenvironment{abstract}{%
    \setlength{\parindent}{0in}%
    \setlength{\parskip}{0in}%
    \bfseries%
    }{\par\vspace{-6pt}}

\newenvironment{addendum}{%
    \setlength{\parindent}{0in}%
    \small%
    \begin{list}{Acknowledgements}{%
        \setlength{\leftmargin}{0in}%
        \setlength{\listparindent}{0in}%
        \setlength{\labelsep}{0em}%
        \setlength{\labelwidth}{0in}%
        \setlength{\itemsep}{12pt}%
        }
    }
    {\end{list}\normalsize}

\renewcommand{\figurename}{\textbf{Figure}}

\newcommand{\suppmv}{Supplementary Movie}
\newcommand{\suppfig}{Supplementary Figure}

\newcommand\ion[2]{#1$\;${\small\rmfamily\rom{#2}}\relax}%
\newcommand*{\rom}[1]{\uppercase\expandafter{\romannumeral #1\relax}}

\newcommand{\aap}{    {\it Astron. Astrophys.}}
\newcommand{\aaps}{   {\it Astron. Astrophys. Suppl.}}
\newcommand{\aapr}{   {\it Astron. Astrophys. Rev.}}

\newcommand{\apj}{    {\it Astrophys. J.}}
\newcommand{\apjl}{   {\it Astrophys. J. Lett.}}

\newcommand{\araa}{   {\it Ann. Rev.  Astron. Astrophys.}}

\newcommand{\jgr}{    {\it J. Geophys. Res.}}

\newcommand{\nat}{    {\it Nature}}

\newcommand{\solphys}{{\it Solar Phys.}}
\newcommand{\ssr}{    {\it Space Sci. Rev.}}

\title{Investigating Energetic X-Shaped Flares on the Outskirts of A Solar Active Region} 
\date{\vspace{-5ex}}
\author{Rui Liu$^{1,2,*}$, Jun Chen$^{1}$, Yuming Wang$^{1,3}$, Kai Liu$^{1,4}$}

\begin{document}

\maketitle

\begin{affiliations}
 \item CAS Key Laboratory of Geospace Environment, Department of Geophysics and Planetary Sciences, University of Science and Technology of China, Hefei 230026, China
 \item Collaborative Innovation Center of Astronautical Science and Technology, Hefei 230026, China
 \item Synergetic Innovation Center of Quantum Information \& Quantum Physics, University of Science and Technology of China, Hefei 230026, China
 \item Mengcheng National Geophysical Observatory, School of 
  	Earth and Space Sciences, University of Science and Technology of China, Hefei 230026, China
\end{affiliations}

\begin{abstract}
Typical solar flares display two quasi-parallel, bright ribbons on the chromosphere. In between is the polarity inversion line (PIL) separating concentrated magnetic fluxes of opposite polarity in active regions (ARs). Intriguingly a series of flares exhibiting X-shaped ribbons occurred at the similar location on the outskirts of NOAA AR 11967, where magnetic fluxes were scattered, yet three of them were alarmingly energetic. The X shape, whose center coincided with hard X-ray emission, was similar in UV/EUV, which cannot be accommodated in the standard flare model. Mapping out magnetic connectivities in potential fields, we found that the X morphology was dictated by the intersection of two quasi-separatrix layers, i.e., a hyperbolic flux tube (HFT), within which a separator connecting a double null was embedded. This topology was not purely local but regulated by fluxes and flows over the whole AR. The nonlinear force-free field model suggested the formation of a current layer at the HFT, where the current dissipation can be mapped to the X-shaped ribbons via field-aligned heat conduction. These results highlight the critical role of HFTs in 3D magnetic reconnection and have important implications for astrophysical and laboratory plasmas.
\end{abstract}

Solar flares are the most powerful events in the solar system. A large flare can release $\sim10^{32}$ erg of energy (the equivalent of over 20 million 100-megaton hydrogen bombs) in tens of minutes, emitting radiation across the entire electromagnetic spectrum and ejecting relativistic particles into interplanetary space, which may jeopardize space-borne and ground-based technological systems \cite{Schwenn2006}. It is well known that such an energy is gradually accumulated in the corona via emergence of magnetic flux through the photosphere and small- and large-scale convective motions in the photospheric layers that shuffle around the footpoints of coronal magnetic field lines \cite{wiegelmann12}. These processes feed magnetic energy and helicity into the solar atmosphere, the latter of which measures the twist and linkage of magnetic field lines \cite{berger84}. There is mounting observational evidence supporting that magnetic reconnection is the key process in the sudden conversion of the accumulated magnetic free energy into thermal and kinetic energies \cite{Priest&Forbes2002}. In particular, the morphology and dynamics of flare ribbons on the chromosphere, which are produced by field-aligned accelerated electrons and heat conduction, provide valuable diagnosis on the reconnection process in the corona \cite{Fletch2011}.

Magnetic reconnection tends to occur at the so-called \emph{structural skeletons} \cite{Titov2007,Titov2009} of magnetic field. Topological skeletons \cite{Longcope2005review}, where the field-line mapping from one footpoint to another is discontinuous, include null points, where the magnetic field vanishes, and separatrix surfaces, which define the boundary of topologically distinct domains. Quasi-skeletons, where the field-line mapping has a steep yet finite gradient, are also known as quasi-separatrix layers (QSLs) \cite{Priest&Demoulin1995}. With magnetic connectivity mapped out by the squashing factor $Q$, QSLs can be defined as high-Q structures, whereas $Q\rightarrow\infty$ at topological skeletons \cite{Titov2002}. This indicates a close relation between QSLs and topological skeletons, and a conversion between them under certain conditions is possible \cite{Restante2009}. Two separatrix surfaces or QSLs may intersect at a separator or a quasi-separator, respectively. A combination of two intersecting QSLs is termed as a hyperbolic flux tube (HFT)\cite{Titov2002}, because of its X-type cross section. Nulls, separators and HFTs are considered to be preferential locations for the concentration of strong currents and the subsequent rapid dissipation \cite{Titov2003,Galsgaard2003,Aulanier2005}. Structural skeletons provide a robust and powerful tool to understand complex observations, and also a solid base for the investigation on dynamics and energetics \cite{Longcope2005review,demoulin06,demoulin07}.

Major flares almost always occur in the neighborhood of PILs separating strong magnetic fields of opposite polarity in ARs. Only a small fraction of flares occur in plages with only small or no sunspot, which are often associated with filament eruptions \cite{Dodson&Hedeman1970,Harvey1986,Rausaria1992,Li1995,Borovik&Myachin2002} and hence fall into the category of the classical ``two-ribbon'' flares, with only a few exceptions in the literature \cite{Sersen&Valnicek1993}. Typically the two ribbons are connected by flare loops across the PIL, and move away from each other as the flare progresses. This can be well explained by the ``standard'' flare model \cite{Carmichael1964, Sturrock1966, Hirayama1974, Kopp1976}. In this two-dimensional model, a rising magnetic flux rope above the PIL stretches the overlying field lines, resulting in a vertical current sheet underneath the rope, where magnetic reconnection occurs successively at an increasingly higher altitude, as a positive feedback is established between the reconnection and the rising of the rope. The two parallel ribbons hence correspond to the footpoints of newly reconnected field lines, because of the assumed symmetry along the PIL. 

Recently, flares exhibiting a circular ribbon have received a lot of attention \cite{Masson2009,Wang&Liu2012,Sun2013,Jiang2014,Vemareddy&Wiegelmann2014,Joshi2015,Liu2015}. These flares cannot be accommodated by the classical two-dimensional model, but can well fit into the fan-spine topology of a three-dimensional null point. The circular ribbon is interpreted as the intersection of a dome-shaped separatrix surface, known as the fan surface of the null, with the chromosphere, while the spine field line threading the fan at the null may anchor at a central ribbon inside the circular ribbon and a remote ribbon outside \cite{Masson2009}. 

Here, we present the rare observations of a series of flares exhibiting X-shaped ribbons (also referred to hereafter as X-shaped flares). These flares did not occur in the neighborhood of internal PILs, but repeatedly in the same place on the outskirts of the active region. However, three of them were alarmingly energetic, ranked as M class. These observations demonstrate that even the outskirts of solar active regions has the potential to produce high-level space weather events, and the X-shaped ribbons highlight the three-dimensional nature of magnetic reconnection in flares. As the flares occurred not far away from the solar disk center, the magnetic field measurements were reliable, providing a precious opportunity to study the magnetic topology as well as the evolution, including creation and destruction, of the relevant structural skeletons. 

\section*{Results}

\subsection*{Instruments and Datasets}
\vskip 5pt
The flares with X-shaped ribbons occurred in NOAA AR 11967, which was monitored by vector magnetograms obtained by the Helioseismic and Magnetic Imager (HMI \cite{hoeksema14}) onboard the Solar Dynamics Observatory (SDO \cite{pesnell12}). The vector magnetograms used in this study were disambiguated and deprojected to the heliographic coordinates with a Lambert (cylindrical equal area; CEA) projection method, resulting in a pixel scale of $0.03^\circ$ (or 0.36 Mm) \cite{Bobra2014}. Note that the Carrington system was adopted so that the longitude of a solar feature remains approximately constant.

The flares produced by AR 11967 were observed with the Atmospheric Imaging Assembly (AIA \cite{lemen12}) onboard SDO. AIA includes seven EUV passbands, i.e., 131~{\AA} (primarily \ion{Fe}{21} for flare plasma, peak response temperature $\log T = 7.05$; primarily \ion{Fe}{8} for ARs, $\log T = 5.6$ \cite{ODwyer2010}), 94~{\AA} (\ion{Fe}{18}, $\log T = 6.85$), 335~{\AA} (\ion{Fe}{16}, $\log T = 6.45$), 211~{\AA} (\ion{Fe}{14}, $\log T = 6.3$), 193~{\AA} (\ion{Fe}{24} for flare plasma, $\log T = 7.25$; \ion{Fe}{12} for ARs, $\log T = 6.2$) and 171~{\AA} (\ion{Fe}{9}, $\log T = 5.85$), 304~{\AA} (\ion{He}{2}, $\log T = 4.7$), and two UV passbands, i.e., 1600~{\AA} (\ion{C}{4}, $\log T = 5.0$) and 1700~{\AA} (continuum). The instrument takes full-disk images with a spatial scale of 0.6 arcsec pixel$^{-1}$ and a cadence of 12 s for EUV and 24 s for UV passbands. Two of the flares (Figure~\ref{fig:flare3} and \suppfig~\ref{suppfig:flare1}) were also observed in hard X-rays (HXRs) by the Reuven Ramaty High-Energy Solar Spectroscopic Imager (RHESSI \cite{lin02}).

It is well known that flare emission in 1600~{\AA} forms the flare ribbons, or the feet of flare loops, since this passband is dominated by \ion{C}{4} line emission, an optically thin line formed at $10^5$ K in the upper chromosphere and transition region. The 1700~{\AA} passband is dominated by UV continuum emission formed at the temperature minimum region (4400--4700 K). Flare enhancement in this passband reflects also footpoint emission, typically attributed to photoionization excited by UV emission from above \cite{Machado&Henoux1982,Doyle&Phillips1992}.

\subsection*{Flare Observation}
\begin{figure}
	\centering
	\includegraphics[width=\hsize]{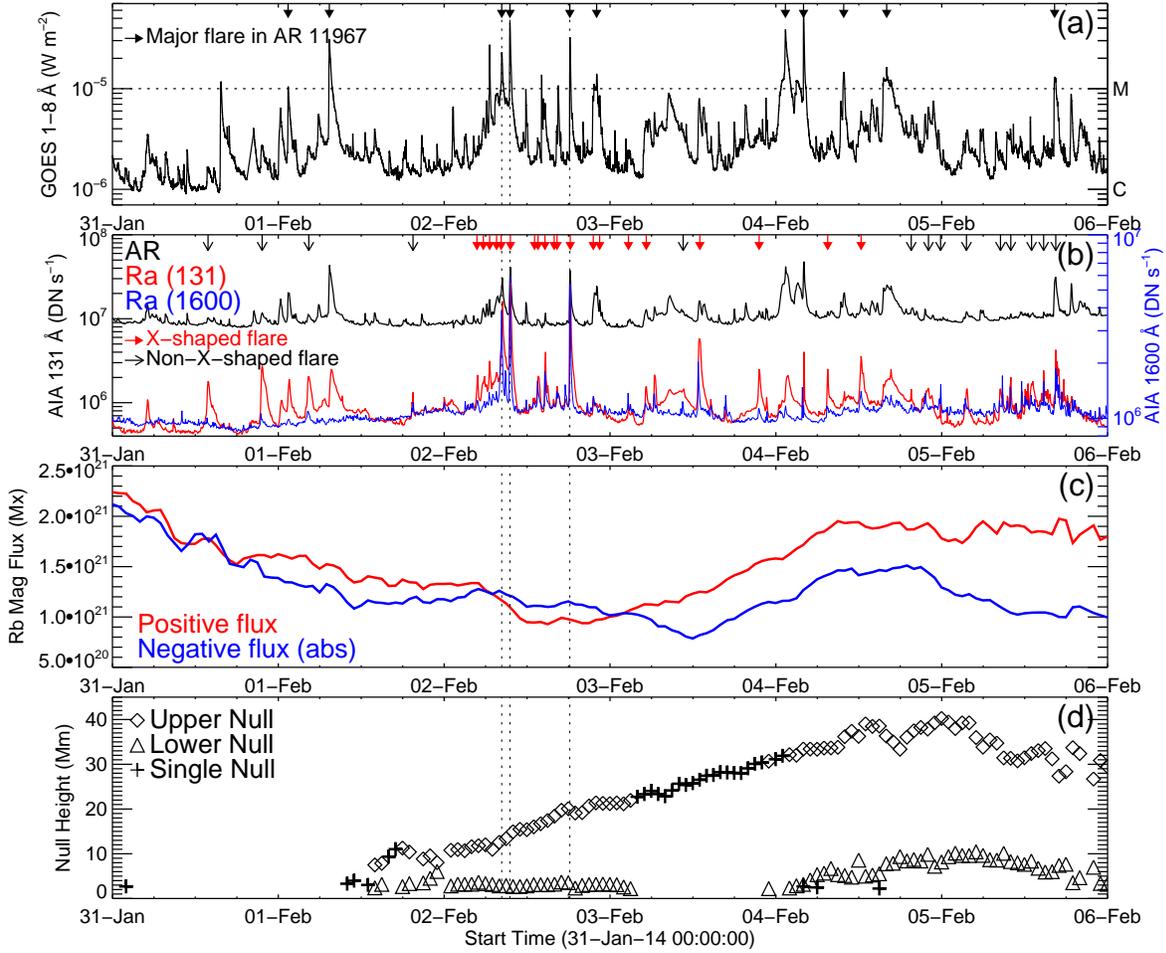}
	\caption{\small Flares in relation to AR 11967. a) GOES 1--8~{\AA} light curve. Arrows on the top mark the peak times of major flares (M-class and above) in this AR from the beginning of January 31 till end of February 5, during which the AR crossed the solar disk from about $45^\circ$E to $45^\circ$W. b) Integrated emission over the whole AR in 131~{\AA} (black) and the regional emission integrated over Ra (rectangle in Figure~\ref{fig:bfield}(b)) in 131~{\AA} (red) and 1600~{\AA} (blue). X-shaped flares are marked by red arrows while non-X-shaped flares by black arrows. c) Magnetic fluxes in Rb (Figure~\ref{fig:bfield}(b)). Positive fluxes and absolute values of negative fluxes are denoted in red and blue, respectively. d) Heights of the detected magnetic null points. The lower null is marked with a triangle and the upper null with a diamond when a double null is present. A single null is marked by a `+' symbol. The peak times of the three XMFs are denoted by vertical dotted lines. \label{fig:ltc} }
\end{figure}

\vskip 5pt
Three flares on 2014 February 2 (Figure~\ref{fig:flare3}, \suppfig s~\ref{suppfig:flare1} and \ref{suppfig:flare2}) were outliers among the 11 major flares (Table~\ref{table}; M-class and above; marked by arrows in Figure~\ref{fig:ltc}(a)) that occurred in AR 11967 from the beginning of 2014 January 31 till the end of February 5, when the AR crossed the solar disk from about $45^\circ$E to $45^\circ$W. These three outliers (peak times marked by vertical lines in Figure~\ref{fig:ltc}) had unique X-shape ribbons and occurred in a facular (or plage) region of AR 11967, where the magnetic field is typically weaker ($\sim100$ G), concentrated in much smaller bundles ($\sim1''$) than in sunspots ($\sim1000$G; $\sim100''$), while all the other non-X-shaped major flares took place near the internal PILs separating sunspots of opposite polarity (cf. Figure~\ref{fig:bfield}(b)). Here the X-shaped major flare (XMF) occurring at 18:11 UT (soft X-ray peak) is taken as an exemplary event (Figure~\ref{fig:flare3}; \suppmv~1), as the other two XMFs taking place at 08:20 and 09:31 UT exhibited almost the same features (\suppfig s~\ref{suppfig:flare1} and \ref{suppfig:flare2} and \suppmv s~6 and 7). Besides the three XMFs, there were many other weaker flaring activities in this region, as shown by the time series of 131 and 1600~{\AA} emission (Figure~\ref{fig:ltc}(b)) integrated over a rectangular region Ra (Figure~\ref{fig:bfield}(b)) centered on the XMFs, using co-registered AIA images at 5-min cadence, as compared to the integrated 131~{\AA} emission over the whole AR. Most of the X-shaped flares in Ra (marked by red solid arrows at the top of Figure~\ref{fig:ltc}(b)) occurred on February 2 and early February 3, while most of non-X-shaped flares in Ra (black hollow arrows) happened before February 2 and after late February 4. One can also see that some of the non-X-shaped flares did not cause significant response in 1600~{\AA}, while for almost all the X-shaped flares the 131~{\AA} and 1600~{\AA} emission in Ra were highly correlated (Figure~\ref{fig:ltc}(b)). There were even weaker activities missing in this approach, such as the event reported by Yang et al. \cite{Yang2015}, which exhibited an X morphology in both H$\alpha$ and EUV and was interpreted in terms of the classic X-type reconnection. All of these X-shaped flares or flaring activities were confined, without being associated with a coronal mass ejection (CME) or a jet. 

It is extraordinary that the flare emission during the impulsive phase had a similar X shape in AIA's all 9 UV/EUV passbands (Figure~\ref{fig:flare3}). Usually the six EUV passbands excluding 304~{\AA} feature flare loops in the corona, whereas 304, 1600 and 1700~{\AA} passbands feature flare ribbons in the chromosphere. The observed X-shaped emission in EUV, however, must originate from the feet of the flare loops due to the similarity with UV ribbons. It was recently noticed that plasma in flare ribbons can be heated to $\sim\,$10 MK during the early impulsive phase before flare loops are filled with evaporated chrompospheric plasma, which indicates intense heating of the lower atmosphere \cite{Fletcher2013,Simoes2015,Polito2016}. The heating mechanism, however, is still under debate. In contrast to UV/EUV emissions, the HXR emission is concentrated at the center of the X shape (Figure~\ref{fig:flare3}), for both the thermal (6--15) and nonthermal (25--50 keV) energy ranges (see \suppfig~\ref{suppfig:spectra}), indicating the projected location where accelerated electrons were interacting with plasma. Two weak, compact nonthermal HXR sources, as compared to the major source at the center of the X shape, were co-located with the most intense UV/EUV patches along the northwestern arm of the X shape. It is clear that this X morphology was not produced by a single classic X-type magnetic reconnection, or reconnections between two crossing, current-carrying loops \cite{Aschwanden1999}, either of which would have produced four flare kernels in UV and/or HXRs. Instead, we observed extended X-shaped flare ribbons in UV/EUV and a compact HXR source at the center of the X shape. The analysis of photospheric magnetic fields will shed light on the nature of the reconnection. 

Superimposing the contours of the line-of-sight (LOS) component of the photospheric magnetic field onto the UV 1700~{\AA} image, one can see that the adjacent endpoints of the X shape were associated with opposite polarities (Figures~\ref{fig:flare3}; \suppfig s~\ref{suppfig:flare1} and \ref{suppfig:flare2}). These four flux concentrations are labeled P1-N1 and p-n. P1 and N1 were associated with two sunspots, whereas p and n with sporadic patches of weak field in the facular region (see also Figure~\ref{fig:bfield}(a)). A third sunspot (labeled P2) was also involved, with the northwestern ``arm'' of the X shape extending to the immediate neighborhood of P2. 

\subsection*{Evolution of Active Region}
\vskip 5pt
Immediately prior to the three XMFs (07:58 UT), the AR was characterized by six major sunspots (Figure~\ref{fig:bfield}(a)), four of them were associated with positive polarity (labeled P1--P4) and two (labeled N1 and N2) with negative polarity. P2 and P4 decayed and became diffused by February 5, while other sunspots were relatively stable. Figure~\ref{fig:bfield}(b--d) show the maps of photospheric flows [km~s$^{-1}$], helicity flux density $h_m$ [Mx$^2$ cm$^{-2}$ s$^{-1}$] and Poynting flux density $p_m$ [erg cm$^{-2}$ s$^{-1}$], which are averaged over a time interval of about 11 hrs covering the XMFs (see Methods). These maps reveal a persistent evolutionary pattern starting in the late January 31 until the mid February 4 (see \suppmv~2), viz., a significant flux of negative polarity emerged to the east of P3, and then migrated eastward into N1 (Figure~\ref{fig:bfield}(b)), forming a channel of negative magnetic fluxes (labeled Nc in Figure~\ref{fig:bfield}(a)) and also a channel of intense negative $h_m$ (Figure~\ref{fig:bfield}(c)) and positive $p_m$ (Figure~\ref{fig:bfield}(d)) between N1 and P3 and to the south of P2. As they approached P1, the flows bifurcated, some were diverted northward, integrated with the moat flows \cite{Solanki2003} of P1 and N1, into the facular region where the three XMFs occurred (olive star symbols in Figure~\ref{fig:bfield}(b)), and others were diverted southeastward along the PIL between P1 and N2. Of the 8 non-X-shaped major flares (see also Table 1), 5 occurred along Nc and 3 along the PIL between P1 and N2. 

It is also remarkable that there was significant flux cancellation ongoing in the facular region to the north of P1-N1, which might be driven by the moat flows of P1 and N1 (see \suppmv~2). Positive and negative fluxes integrated in an irregular region Rb (Figure~\ref{fig:bfield}(b)) are shown in Figure~\ref{fig:ltc}(c). Rb was so selected as to avoid the sunspots P1, N1, and P2. The cancellation started from the beginning of January 31 until mid February 3. After that it appeared to be dominated by the emergence of positive fluxes in Rb and the migration of negative fluxes from N1 into Rb with these flux elements being detached from N1. Studies have argued for the association between flux cancellation and magnetic reconnection \cite{Zwaan1987}.

\subsection*{Magnetic Topology}
\vskip 5pt
To understand the magnetic connectivities in AR 11967, we performed the extrapolation of the coronal potential field on an hourly cadence from January 31 to February 5 (see Methods). Potential field maintains the basic topology although it can only be regarded as a zero-order approximation of the real coronal field. The robustness of structural skeletons has been demonstrated by earlier studies employing various coronal field models \cite{demoulin06,demoulin07,Liu2014}, and also in the present study by switching on and off a ``pre-processing'' procedure on the photospheric boundary (see Methods). However, current-carrying structures, e.g., twisted flux tubes, cannot be recovered in the potential field, and the assumption adopted by the null-locating algorithm \cite{Parnell1996,Haynes&Parnell2007} (see also Methods), that the magnetic field approaches zero linearly toward nulls, might break down with the presence of strong currents. 

For each potential field, we sought for magnetic null points within a box volume, whose bottom was the rectangular region indicated in Figure~\ref{fig:bfield}(a). Most of the identified nulls were located in the facular region to the north of P1 and N1 (see Methods and \suppfig s~\ref{suppfig:nullxy} and \ref{suppfig:nullz}), and sometimes an interesting double-coronal-null configuration was present, i.e., two nulls were separated in heights but have similar X-Y positions. One can see from Figure~\ref{fig:ltc}(d) that double nulls started to appear in late February 1, disappeared in early February 3, and reappeared in the middle of February 4. It is noteworthy that most of the X-shaped flares occurred when the double-null configuration was present and when there was significant flux cancellation in the facular region of interest (Figure~\ref{fig:ltc}). Field lines traced in the neighborhood of the double null just prior to the XMF at 18:11 UT on 2014 February 2 are shown in Figure~\ref{fig:null}(a, e and f) with fan (spine) field lines in magenta (blue).  The upper null was located at $Z=20.08$ Mm and the lower null at $Z=3.55$ Mm (marked by green crosses in Figure~\ref{fig:null}(e and f)) .

We also calculated the squashing factor $Q$ in the same box volume (see Methods). The isosurfaces of $\log_{10}Q=5$ (Figure~\ref{fig:qsl}) consist of two major intersecting QSLs, labeled `S1' and `S2', whose footprints on the photosphere correspond to the high-$Q$ lines in Figure~\ref{fig:null}(a and b). The intersection of S1 and S2 is a quasi-separator by definition. It is well defined immediately above the lower null, as demonstrated by the X-shaped morphology in horizontal cutting planes of $\log_{10}Q$ (Figure~\ref{fig:null}(c and d); see also \suppmv~3), the characteristic of an HFT. The center of X was visually determined and plotted as black dots in Figure~\ref{fig:null}(e and f), showing that the two nulls are located along the quasi-separator. Viewing the QSLs from different perspectives (Figure~\ref{fig:qsl} and \suppmv~4), one can see that the fan (spine) field lines of the upper null, partially visible through the isosurfaces, are embedded within S1 (S2), while the reverse is true for the lower null. The fan plane of the upper null is hence bounded by the spine of the lower null from below within S1, while the fan plane of the lower null is bounded by the spine of the upper null from above within S2. The intersection of the two fan planes hence determines a separator connecting the two nulls \cite{Priest2014book}, which in our case is embedded within the quasi-separator. The quasi-separator extends beyond the upper null, up to as high as about 60 Mm, inclining southward above the upper null and northward below it (Figure~\ref{fig:qsl}(b and c) and \suppmv~4). With or without the nulls, the overall configuration of the HFT remained almost the same as in Figure~\ref{fig:qsl} (see also \suppfig~\ref{suppfig:qsl}) during the time period investigated, due presumably to the relatively slow evolution of AR 11967 (see \suppmv~2). Apparently, the recurrence of X-shaped flares in this region owed its origin to the relatively stable presence of this HFT. 


The double null's fans and spines yield an X shape, but its mismatch with the flare ribbons is not negligible (Figure~\ref{fig:null}(a)). In particular, the southwestern arm of the X shape is left unaccounted for, as the upper null's fan is bounded by the lower null's spine from below. Hence, the double null and the separator cannot be solely responsible for the observed flare ribbons. On the other hand, it was found that the X-shaped ribbons are better matched by high-Q lines on the photosphere (Figure~\ref{fig:null}(a); see also Methods), which delineate the footprints of S1 and S2. The flare ribbons are hence suggested to be mainly energized by the magnetic reconnection at the HFT. 

However, to produce the flare, the configuration must carry a substantial electric current. We built a nonlinear force-free field (NLFFF) with the ``weighted optimization'' method to model the coronal field \cite{wiegelmann04,wiegelmann12}. The vector magnetogram at 17:58 UT was ``pre-processed'' \cite{wiegelmann06} to minimize the net force and torque in the observed photospheric field before being taken as the boundary. By obtaining the squashing factor $Q$ for this NLFFF in the same box volume as above, we found that the basic topology can still be interpreted as two intersecting QSLs, yet the intersection was no longer a 3D curve as in the potential field, but a very complex structure (Figure~\ref{fig:nlff}(d) and \suppmv~5), with the concentration of electric currents (Figure~\ref{fig:nlff}(c)), yet with no null being identified numerically. However, the overall morphology of the intersection retains the X shape, the basic characteristic of an HFT. We hence determined that the HFT in the NLFFF is equivalent in topology to, but much more complex in geometry than, its counterpart in the potential field. Its footprints at the photosphere also compare favorably with the X-shaped flare ribbons (Figure~\ref{fig:nlff}(a)), which strengthens our conclusion that the flare ribbons are closely associated with the HFT. Similar results were found for the other two XMFs.
\section*{Discussion}
\vskip 5pt
A comparison with an idealized quadrupole field consisting of two oppositely directed dipoles with different magnitudes (see Methods) may shed light on the observed topology, which appeared to involve mostly a local quadrupole field as represented by P1-N1 and p-n. An HFT inclining towards the weaker dipole was identified at the intersection of two high-Q surfaces in this idealized quadrupole field (\suppfig s~\ref{suppfig:bquad}--\ref{suppfig:asym}). When the two dipoles are strictly anti-parallel, a null line was found to be coincident with the HFT. Each point along the null line is associated with two spines, whose footpoints collectively match the X-shaped footprint of the HFT. The HFT persists while the null line perishes when the two dipoles are not perfectly anti-parallel. However, the arms of this X shape diverge toward the stronger dipole and converge toward the weaker one, and the strength of the two dipoles has to be comparable to yield a symmetric X shape. In contrast, the observed flare ribbons were more extended at the weak field side p-n than at the strong side P1-N1 (Figure~\ref{fig:flare3} and \suppfig s~\ref{suppfig:flare1} and \ref{suppfig:flare2}). The separator in observation can be regarded as a generalized form of the null line and did incline toward the weak field as the null line, but above the upper null, the quasi-separator inclined backwards toward the strong field. This suggests that the production of the X-shaped flare ribbons was not just dictated by the local quadrupole field, but must be regulated by all the other flux concentrations as well as photospheric flows within the AR.

The above analysis converges to a self-consistent physical scenario: the flux emergence and flows associated with the channel Nc energize the AR by injecting magnetic helicity and Poynting fluxes into the solar atmosphere, which results in the accumulation of magnetic free energy and stress in the neighborhood of the internal PILs, as well as in the facular region of interest, above which the HFT is probably pinched into a current layer, as suggested by the NLFFF model as well as studies with a broad variety of photospheric motions applied to its feet \cite{Titov2003,Galsgaard2003,Aulanier2005}. When magnetic reconnections occur at this current layer, the field-aligned heat conduction channels out the energy released at the HFT and produces the UV/EUV flare ribbons residing at the high Q-lines that delineate the HFT footprints. In contrast, the HXR emission was concentrated at the center of the X shape, which suggests that most of the HXR-producing electrons were thermalized at the HFT, with only a fraction of them leaking out to produce sporadic HXR emission at the flare ribbons (Figure~\ref{fig:flare3} and \suppfig s~\ref{suppfig:flare1} and \ref{suppfig:spectra}). How the electrons were accelerated and trapped at the HFT is beyond the scope of this study, but certainly an important area to pursue for future research. The homologousness of these X-shaped flares can be attributed to the repetitive energization and relaxation of the HFT, which proved to be persistent during the time period investigated. These results provide insight into the nature of 3D magnetic reconnection in solar flares, which could be constructive in the efforts of flare forecasting, and may also have important implications for other astrophysical as well as laboratory plasmas. 

\section*{Methods}
\subsection*{Helicity and Poynting Flux} \label{append:hp}
\vskip 5pt
The flow field in Figure~\ref{fig:bfield}(b) was obtained by applying the Differential Affine Velocity Estimator for Vector Magnetograms (DAVE4VM) \cite{schuck08} to the time-series of deprojected, registered vector magnetograms. The window size used in DAVE4VM was chosen to be 19 pixels, following Liu et al. \cite{liuy14}. 
 
With the flow field, we were able to calculate the relative helicity flux across its photospheric boundary $S$ with the following formula \cite{berger84,liuy12,liuy14}:
\begin{equation}
\left.\frac{dH}{dt}\right|_S=2\int_S(\mathbf{A}_p\cdot\mathbf{B}_t)V_{\perp n}\,dS-2\int_S(\mathbf{A}_p\cdot\mathbf{V}_{\perp t})B_n\,dS, \label{eq:helicity}
\end{equation}
where $\mathbf{A}_p$ is the vector potential of the reference potential field $\mathbf{B}_p$ that shares the same $B_z$ on the photospheric boundary as the observation; $t$ and $n$ refer to the tangential and normal directions, respectively. $\mathbf{V}_\perp$ is the photospheric velocity perpendicular to magnetic field lines, which is obtained by subtracting the field-aligned plasma flow $(\mathbf{V}\cdot\mathbf{B})\mathbf{B}/B^2$ from the velocity derived by DAVE4VM \cite{liuy12}. The two terms on the right hand side of Eq.~\ref{eq:helicity} describe helicity injection due to flux emergence ($V_{\perp n}$) and photospheric motions ($\mathbf{V}_{\perp t}$) that shear and braid field lines, respectively. Similarly, both flux emergence and tangential motions contribute to the Poynting flux across the photospheric boundary \cite{kusano02}, 
\begin{equation}
\left.\frac{dP}{dt}\right|_S=\frac{1}{4\pi}\int_S B_t^2V_{\perp n}\,dS-\frac{1}{4\pi}\int_S(\mathbf{B}_t\cdot\mathbf{V}_{\perp t})B_n\,dS. \label{eq:poynting}
\end{equation}
Here our error analysis follows Appendix B in Liu et al. \cite{Liu2016}. The typical errors of helicity and Poynting fluxes are of order $10^{36}$ Mx$^2$~s$^{-1}$ and $10^{25}$ erg~s$^{-1}$, respectively, less than 1 part in 10. For AR 11967 as a whole (\suppfig~\ref{suppfig:lc}), we found no significant injection of magnetic flux [Mx], helicity flux [Mx$^2$ s$^{-1}$], or Poynting flux [erg s$^{-1}$] during the time period of interest. An injection of helicity and Poynting fluxes started from the early February 5, but apparently it was not related with any major flares. On the other hand, examining the flux density maps in detail (Figure~\ref{fig:bfield}(c--d)), we found a significant local injection of negative helicity flux and positive Poynting flux, associated with the eastward flows in the channel Nc. 

\subsection*{Extrapolation of Coronal Potential Field} \label{append:bfield}
\vskip 5pt
We calculated the potential field with both the Fourier transformation \cite{Alissandrakis1981} and the Green function method. Both the original and a ``pre-processed'' $B_z$ was adopted as the photospheric boundary of the extrapolation. The aim of the pre-processing, as applied to the $2\times2$ rebinned Space-Weather HMI Active Region Patches data, was to make the vector magnetograms best suit the force-free condition \cite{wiegelmann06}. 

The calculation was carried out within a box of $720\times 344\times 256$ uniformly spaced grid points, corresponding to $524.5\times250.6\times186.5$ Mm, whose photospheric FOV spanned [86.6, 129.8] in Carrington longitudes, and [-23.4, -2.7] in latitudes. Magnetic flux was roughly balanced within the FOV, the ratio between positive and (absolute) negative flux was 1.07 at the beginning of January 31, increasing more or less monotonously toward 1.36 by the end of February 5. The results shown in Figures~\ref{fig:null} and \ref{fig:qsl} were given by the potential field constructed with the Fourier method and the pre-processed boundary. Other potential fields yielded very similar results as far as magnetic topology is concerned (\suppfig s~\ref{suppfig:nullxy} and \ref{suppfig:nullz}).   

\subsection*{Map of Magnetic Connectivities} \label{append:qfactor}
\vskip 5pt
We investigated magnetic connectivities within the extrapolated field by tracing field lines pointwise with a fourth-order Runge-Kutta method to ensure high precision. The mapped footpoints of field lines were used to calculate the squashing factor $Q$ of elemental magnetic flux tubes \cite{Titov2002}. Basically, for a mapping through the two footpoints of a field line $\Pi_{12}:  \mathbf{r}_1(x_1,\ y_1)\mapsto \mathbf{r}_2(x_2,\ y_2)$, the squashing factor $Q$ associated with the field line is \cite{Titov2002}
\begin{equation} \label{eq:q}
Q\equiv \frac{a^2+b^2+c^2+d^2}{|\mathrm{det}\,D|},
\end{equation}
where $a,b,c,d$ are elements of the Jacobian matrix 
\begin{equation} \label{eq:d12} 
D=\left[\frac{\partial \mathbf{r}_2}{\partial \mathbf{r}_1}\right]= 
\begin{pmatrix}
\displaystyle \frac{\partial x_2}{\partial x_1} & \displaystyle \frac{\partial x_2}{\partial y_1} \\
\displaystyle \frac{\partial y_2}{\partial x_1} & \displaystyle \frac{\partial y_2}{\partial y_1}
\end{pmatrix} \equiv 
\begin{pmatrix}
a & b \\
c & d
\end{pmatrix}.
\end{equation}
Quasi-separatrix layers (QSLs) as defined by high-$Q$ values are often complex three-dimensional structures, hence their visualization can be facilitated by calculating $Q$ in a 3D volume box. This was done by stacking up $Q$-maps in uniformly spaced cutting planes \cite{Liu2016}. In Figure~\ref{fig:qsl}, the original grids were refined by 4 times in the calculation of $Q$. 

\subsection*{Identification of Null Points} \label{append:null}
\vskip 5pt
To locate null points in a 3D magnetic field with uniform grids, we assume that within each cell the field is linear or trilinear, following Haynes \& Parnell \cite{Haynes&Parnell2007}. Under this assumption, a cell is excluded in the further analysis if any of the three magnetic field components $[B_x,\,B_y,\,B_z]$ is nonzero and of the same sign at each corner of the cell. More cells can be excluded by checking whether the intersection curves of the isosurfaces $B_x=0$, $B_y=0$, and $B_z=0$ pierce the cell faces in pair \cite{Haynes&Parnell2007}. Two methods are used to locate the null position to subgrid precision. One may conduct a brute-force search for $\min B$ in each candidate cell. First the cell is uniformly divided into (say 1000) miniature cells. The miniature cell centered on the grid point with $\min B$ is further divided and a new $\min B$ is located. This is done recursively until the number of recursion exceeds, say, 3 times, or, $\min B < 10^{-4}$ G. A null point is located if the latter condition is satisfied.  Alternatively, one may solve $\mathbf{B}=0$ with the iterative Newton-Raphson method \cite{Haynes&Parnell2007}. We confirmed that null points detected in both methods are consistent within the limits of numerical accuracy. However, the Newton-Raphson method is generally more efficient and has the potential to identify multiple nulls within a candidate cell, e.g., in the case of a null line (see below), if one initiates the iteration from a sufficiently large number of subgrid points. 

The local magnetic structure in the vicinity of a null point is determined by the tensor $M_{ij}=\partial B_i/\partial x_j$ \cite{Parnell1996}. The structure is relatively simple for potential fields, since without currents $\mathbf{M}$ is symmetric and has three real eigenvalues, whose sum is zero. For an isolated null, there is one eigenvalue of opposite sign to the other two, and its associated eigenvector determines the spine while the other two eigenvectors define the fan plane. This differs for a non-isolated null, as exemplified by a null line in an idealized quadrupole field (see below). 

There are subtle differences for nulls identified in the potential field constructed with different methods (Green function vs. Fourier transform) and different boundary conditions (original vs. pre-processsed), but the time period when the double-null configuration was present and the locations of these nulls are essentially the same (\suppfig s~\ref{suppfig:nullxy} and \ref{suppfig:nullz}). This demonstrates the robustness of such topological skeletons.

\subsection*{Idealized Quadruple Field} \label{append:nullline}
\vskip 5pt
Here an idealized quadrupole field, similar to Sun et al. \cite{Sun2013}, was employed to shed light on the magnetic topology involved in the X-shaped flares. For simplicity, two dipoles were placed in an opposite direction below a Cartesian computation domain
$(-255.5\leq x, y \leq 255.5,\,0\leq z\leq 511)$: $\mathbf{m}_1 = 1.024 \times 10^9 \times (-1,\,0,\,0)^T$ at $\mathbf{d}_1 = (0,\,-60,\,-80)^T$ , and $\mathbf{m}_2 = \alpha\times 1.024\times10^9 \times (1,\,0,\,0)^T$ at $\mathbf{d}_2 = (0,\,60,\,-80)^T$, where $\alpha$ is a fraction number between 0 and 1. The sum of the two dipole fields yields a potential field with a magnitude of 2000 G at the photosphere ($z=0$),
\begin{equation}
\mathbf{B(r)}=\sum_{i=1}^2\frac{3(\mathbf{m}_i\cdot\mathbf{r}_i)\mathbf{r}_i}{r_i^5} - \frac{\mathbf{m}_i}{r_i^3}, \label{eq:bquad}
\end{equation}
where $\mathbf{r}_i=\mathbf{r}-\mathbf{d}_i$. With this setup, $\mathbf{m}_i\cdot\mathbf{r}_i=0$ and $B_y=B_z=0$ at the plane  $x=0$. A null line along which $\mathbf{B}=0$ was found to be a circular curve within this plane of symmetry, i.e.,
\begin{equation*}
|\mathbf{r}-\mathbf{C(\mathbf{\alpha, d_1},\mathbf{d_2})}|=|\mathbf{R(\alpha,\mathbf{d_1},\mathbf{d_2})}|,
\end{equation*}
where 
\begin{equation*}
\mathbf{C}=\frac{\mathbf{d_2}-\alpha^{2/3} \mathbf{d_1}}{1-\alpha^{2/3}} \quad\text{and}\quad \mathbf{R}=\frac{\alpha^{1/3}}{1-\alpha^{2/3}}(\mathbf{d_2}-\mathbf{d_1}).
\end{equation*}
When $\alpha=1$, however, it degenerates into a line within the same plane, 
\begin{equation*}
(\mathbf{d_2}-\mathbf{d_1})\cdot\mathbf{r}+\frac{1}{2}(d_2^2-d_1^2)=0. 
\end{equation*}
We have tested our null-locating code with this quadrupole field and found that all the nulls identified are located along the analytical null line within the limits of numerical accuracy. 

Since $\mathbf{B}$ is potential, we again have three real eigenvalues for $\mathbf{M}=[\partial\mathbf{B}/\partial\mathbf{r}]$ given an arbitrary point on the null line, but one of the eigenvalues is precisely zero, say, $\lambda_1>0,\,\lambda_2<0,\,\lambda_3=0$, and $\lambda_1+\lambda_2=0$. Formally, a position vector $\mathbf{r}$ on a field line near the null can be written as \cite{Parnell1996}
\begin{equation}
\mathbf{r}(k) = c_1e^{\lambda_1k}\mathbf{v}_1 + c_2e^{\lambda_2k}\mathbf{v}_2 + c_3e^{\lambda_3k}\mathbf{v}_3, \label{eq:r}
\end{equation}
where $\mathbf{v}_i$ are eigenvectors, $k$ is an arbitrary parameter and $c_i$ are constant along the field line, whose behavior near the null is determined as follows,
\begin{equation}
\frac{d\mathbf{r}(k)}{dk} = \mathbf{M}\cdot\mathbf{r}(k).
\end{equation}
Hence, when one traces a field line forward ($k\rightarrow\infty$) or backward ($k\rightarrow -\infty$) away from the null, the $\mathbf{v}_3$ term in Eq.~\ref{eq:r} becomes irrelevant, and all the field lines are either directed away from the null along $\mathbf{v}_1$ or toward the null along $\mathbf{v}_2$. Thus, each point on the null line owns two spines. \suppfig~\ref{suppfig:bquad} gives three examples with $\alpha = [0.2,0.5,0.8]$. One can see that the null line inclines toward the weaker dipole and that the footpoints of all the spines collectively produce an X shape, if the magnitudes of the two dipoles are comparable.

Mapping out the magnetic connectivities (\suppfig~\ref{suppfig:toytop}), one can see that the null line corresponds to the intersection of two high-Q surfaces, whose footprints are coincident with the footpoints of the spines. This again demonstrates the intimate relation between topological skeletons and QSLs \cite{Titov2002,Restante2009}. However, a slight rotation of $\mathbf{m}_2$ about the $z$-axis breaks the symmetry and the null line disappears completely. Only one single null is identified, when the rotational angle is within about $\pm20^\circ$, beyond which no null is found within the computational domain. In contrast, the HFT structure is stable; it again leans toward the weaker dipole, and its footprint leaves a skewed X shape on the surface when $\alpha$ approaches unit (\suppfig~9). 


\begin{thebibliography}{10}
\expandafter\ifx\csname url\endcsname\relax
  \def\url#1{\texttt{#1}}\fi
\expandafter\ifx\csname urlprefix\endcsname\relax\def\urlprefix{URL }\fi
\providecommand{\bibinfo}[2]{#2}
\providecommand{\eprint}[2][]{\url{#2}}

\bibitem{Schwenn2006}
\bibinfo{author}{{Schwenn}, R.}
\newblock \bibinfo{title}{{Space Weather: The Solar Perspective}}.
\newblock \emph{\bibinfo{journal}{Living Reviews in Solar Physics}}
  \textbf{\bibinfo{volume}{3}} (\bibinfo{year}{2006}).

\bibitem{wiegelmann12}
\bibinfo{author}{{Wiegelmann}, T.} \emph{et~al.}
\newblock \bibinfo{title}{{How Should One Optimize Nonlinear Force-Free Coronal
  Magnetic Field Extrapolations from SDO/HMI Vector Magnetograms?}}
\newblock \emph{\bibinfo{journal}{\solphys}} \textbf{\bibinfo{volume}{281}},
  \bibinfo{pages}{37--51} (\bibinfo{year}{2012}).

\bibitem{berger84}
\bibinfo{author}{{Berger}, M.~A.}
\newblock \bibinfo{title}{{Rigorous new limits on magnetic helicity dissipation
  in the solar corona}}.
\newblock \emph{\bibinfo{journal}{Geophysical and Astrophysical Fluid
  Dynamics}} \textbf{\bibinfo{volume}{30}}, \bibinfo{pages}{79--104}
  (\bibinfo{year}{1984}).

\bibitem{Priest&Forbes2002}
\bibinfo{author}{{Priest}, E.~R.} \& \bibinfo{author}{{Forbes}, T.~G.}
\newblock \bibinfo{title}{{The magnetic nature of solar flares}}.
\newblock \emph{\bibinfo{journal}{\aapr}} \textbf{\bibinfo{volume}{10}},
  \bibinfo{pages}{313--377} (\bibinfo{year}{2002}).

\bibitem{Fletch2011}
\bibinfo{author}{{Fletcher}, L.} \emph{et~al.}
\newblock \bibinfo{title}{{An Observational Overview of Solar Flares}}.
\newblock \emph{\bibinfo{journal}{\ssr}} \textbf{\bibinfo{volume}{159}},
  \bibinfo{pages}{19--106} (\bibinfo{year}{2011}).

\bibitem{Titov2007}
\bibinfo{author}{{Titov}, V.~S.}
\newblock \bibinfo{title}{{Generalized Squashing Factors for Covariant
  Description of Magnetic Connectivity in the Solar Corona}}.
\newblock \emph{\bibinfo{journal}{\apj}} \textbf{\bibinfo{volume}{660}},
  \bibinfo{pages}{863--873} (\bibinfo{year}{2007}).

\bibitem{Titov2009}
\bibinfo{author}{{Titov}, V.~S.}, \bibinfo{author}{{Forbes}, T.~G.},
  \bibinfo{author}{{Priest}, E.~R.}, \bibinfo{author}{{Miki{\'c}}, Z.} \&
  \bibinfo{author}{{Linker}, J.~A.}
\newblock \bibinfo{title}{{Slip-Squashing Factors as a Measure of
  Three-Dimensional Magnetic Reconnection}}.
\newblock \emph{\bibinfo{journal}{\apj}} \textbf{\bibinfo{volume}{693}},
  \bibinfo{pages}{1029--1044} (\bibinfo{year}{2009}).

\bibitem{Longcope2005review}
\bibinfo{author}{{Longcope}, D.~W.}
\newblock \bibinfo{title}{{Topological Methods for the Analysis of Solar
  Magnetic Fields}}.
\newblock \emph{\bibinfo{journal}{Living Reviews in Solar Physics}}
  \textbf{\bibinfo{volume}{2}} (\bibinfo{year}{2005}).

\bibitem{Priest&Demoulin1995}
\bibinfo{author}{{Priest}, E.~R.} \& \bibinfo{author}{{D{\'e}moulin}, P.}
\newblock \bibinfo{title}{{Three-dimensional magnetic reconnection without null
  points. 1. Basic theory of magnetic flipping}}.
\newblock \emph{\bibinfo{journal}{\jgr}} \textbf{\bibinfo{volume}{100}},
  \bibinfo{pages}{23443--23464} (\bibinfo{year}{1995}).

\bibitem{Titov2002}
\bibinfo{author}{{Titov}, V.~S.}, \bibinfo{author}{{Hornig}, G.} \&
  \bibinfo{author}{{D{\'e}moulin}, P.}
\newblock \bibinfo{title}{{Theory of magnetic connectivity in the solar
  corona}}.
\newblock \emph{\bibinfo{journal}{Journal of Geophysical Research (Space
  Physics)}} \textbf{\bibinfo{volume}{107}}, \bibinfo{pages}{1164}
  (\bibinfo{year}{2002}).

\bibitem{Restante2009}
\bibinfo{author}{{Restante}, A.~L.}, \bibinfo{author}{{Aulanier}, G.} \&
  \bibinfo{author}{{Parnell}, C.~E.}
\newblock \bibinfo{title}{{How skeletons turn into quasi-separatrix layers in
  source models}}.
\newblock \emph{\bibinfo{journal}{\aap}} \textbf{\bibinfo{volume}{508}},
  \bibinfo{pages}{433--443} (\bibinfo{year}{2009}).

\bibitem{Titov2003}
\bibinfo{author}{{Titov}, V.~S.}, \bibinfo{author}{{Galsgaard}, K.} \&
  \bibinfo{author}{{Neukirch}, T.}
\newblock \bibinfo{title}{{Magnetic Pinching of Hyperbolic Flux Tubes. I. Basic
  Estimations}}.
\newblock \emph{\bibinfo{journal}{\apj}} \textbf{\bibinfo{volume}{582}},
  \bibinfo{pages}{1172--1189} (\bibinfo{year}{2003}).

\bibitem{Galsgaard2003}
\bibinfo{author}{{Galsgaard}, K.}, \bibinfo{author}{{Titov}, V.~S.} \&
  \bibinfo{author}{{Neukirch}, T.}
\newblock \bibinfo{title}{{Magnetic Pinching of Hyperbolic Flux Tubes. II.
  Dynamic Numerical Model}}.
\newblock \emph{\bibinfo{journal}{\apj}} \textbf{\bibinfo{volume}{595}},
  \bibinfo{pages}{506--516} (\bibinfo{year}{2003}).

\bibitem{Aulanier2005}
\bibinfo{author}{{Aulanier}, G.}, \bibinfo{author}{{Pariat}, E.} \&
  \bibinfo{author}{{D{\'e}moulin}, P.}
\newblock \bibinfo{title}{{Current sheet formation in quasi-separatrix layers
  and hyperbolic flux tubes}}.
\newblock \emph{\bibinfo{journal}{\aap}} \textbf{\bibinfo{volume}{444}},
  \bibinfo{pages}{961--976} (\bibinfo{year}{2005}).

\bibitem{demoulin06}
\bibinfo{author}{{D{\'e}moulin}, P.}
\newblock \bibinfo{title}{{Extending the concept of separatrices to QSLs for
  magnetic reconnection}}.
\newblock \emph{\bibinfo{journal}{Advances in Space Research}}
  \textbf{\bibinfo{volume}{37}}, \bibinfo{pages}{1269--1282}
  (\bibinfo{year}{2006}).

\bibitem{demoulin07}
\bibinfo{author}{{D{\'e}moulin}, P.}
\newblock \bibinfo{title}{{Where will efficient energy release occur in 3-D
  magnetic configurations?}}
\newblock \emph{\bibinfo{journal}{Advances in Space Research}}
  \textbf{\bibinfo{volume}{39}}, \bibinfo{pages}{1367--1377}
  (\bibinfo{year}{2007}).

\bibitem{Dodson&Hedeman1970}
\bibinfo{author}{{Dodson}, H.~W.} \& \bibinfo{author}{{Hedeman}, E.~R.}
\newblock \bibinfo{title}{{Major H{$\alpha$} Flares in Centers of Activity with
  very Small or no Spots}}.
\newblock \emph{\bibinfo{journal}{\solphys}} \textbf{\bibinfo{volume}{13}},
  \bibinfo{pages}{401--419} (\bibinfo{year}{1970}).

\bibitem{Harvey1986}
\bibinfo{author}{{Harvey}, K.}, \bibinfo{author}{{Sheeley}, N., Jr.} \&
  \bibinfo{author}{{Harvey}, J.}
\newblock \bibinfo{title}{{HE I 10830 A Observations of Two-Ribbon Flare-Like
  Events Associated with Filament Disappearances}}.
\newblock In \bibinfo{editor}{{Simon}, P.~A.}, \bibinfo{editor}{{Heckman}, G.}
  \& \bibinfo{editor}{{Shea}, M.~A.} (eds.)
  \emph{\bibinfo{booktitle}{Solar-Terrestrial Predictions}},
  \bibinfo{pages}{198} (\bibinfo{year}{1986}).

\bibitem{Rausaria1992}
\bibinfo{author}{{Rausaria}, R.~R.}, \bibinfo{author}{{Aleem}, S.~M.} \&
  \bibinfo{author}{{Sundara Raman}, K.}
\newblock \bibinfo{title}{{On the triggering of a spotless double-ribbon
  flare}}.
\newblock \emph{\bibinfo{journal}{\solphys}} \textbf{\bibinfo{volume}{142}},
  \bibinfo{pages}{131--141} (\bibinfo{year}{1992}).

\bibitem{Li1995}
\bibinfo{author}{{Li}, K.~J.}, \bibinfo{author}{{Zhong}, S.~H.},
  \bibinfo{author}{{Ding}, Y.~J.}, \bibinfo{author}{{Bai}, J.~M.} \&
  \bibinfo{author}{{Li}, Q.~Y.}
\newblock \bibinfo{title}{{The sunspotless flare on April 12, 1991 and the
  evolution of the neighboring filaments.}}
\newblock \emph{\bibinfo{journal}{\aaps}} \textbf{\bibinfo{volume}{109}}
  (\bibinfo{year}{1995}).

\bibitem{Borovik&Myachin2002}
\bibinfo{author}{{Borovik}, A.~V.} \& \bibinfo{author}{{Myachin}, D.~Y.}
\newblock \bibinfo{title}{{The spotless flare of 16 March 1981 - I. Pre-Flare
  Activations of the Chromospheric Fine Structure}}.
\newblock \emph{\bibinfo{journal}{\solphys}} \textbf{\bibinfo{volume}{205}},
  \bibinfo{pages}{105--116} (\bibinfo{year}{2002}).

\bibitem{Sersen&Valnicek1993}
\bibinfo{author}{{Sersen}, M.} \& \bibinfo{author}{{Valnicek}, B.}
\newblock \bibinfo{title}{{Spotless flares and type II radio bursts}}.
\newblock \emph{\bibinfo{journal}{\solphys}} \textbf{\bibinfo{volume}{145}},
  \bibinfo{pages}{339--345} (\bibinfo{year}{1993}).

\bibitem{Carmichael1964}
\bibinfo{author}{{Carmichael}, H.}
\newblock \bibinfo{title}{{A Process for Flares}}.
\newblock \emph{\bibinfo{journal}{NASA Special Publication}}
  \textbf{\bibinfo{volume}{50}}, \bibinfo{pages}{451} (\bibinfo{year}{1964}).

\bibitem{Sturrock1966}
\bibinfo{author}{{Sturrock}, P.~A.}
\newblock \bibinfo{title}{{Model of the High-Energy Phase of Solar Flares}}.
\newblock \emph{\bibinfo{journal}{\nat}} \textbf{\bibinfo{volume}{211}},
  \bibinfo{pages}{695--697} (\bibinfo{year}{1966}).

\bibitem{Hirayama1974}
\bibinfo{author}{{Hirayama}, T.}
\newblock \bibinfo{title}{{Theoretical Model of Flares and Prominences. I:
  Evaporating Flare Model}}.
\newblock \emph{\bibinfo{journal}{\solphys}} \textbf{\bibinfo{volume}{34}},
  \bibinfo{pages}{323--338} (\bibinfo{year}{1974}).

\bibitem{Kopp1976}
\bibinfo{author}{{Kopp}, R.~A.} \& \bibinfo{author}{{Pneuman}, G.~W.}
\newblock \bibinfo{title}{{Magnetic reconnection in the corona and the loop
  prominence phenomenon}}.
\newblock \emph{\bibinfo{journal}{\solphys}} \textbf{\bibinfo{volume}{50}},
  \bibinfo{pages}{85--98} (\bibinfo{year}{1976}).

\bibitem{Masson2009}
\bibinfo{author}{{Masson}, S.}, \bibinfo{author}{{Pariat}, E.},
  \bibinfo{author}{{Aulanier}, G.} \& \bibinfo{author}{{Schrijver}, C.~J.}
\newblock \bibinfo{title}{{The Nature of Flare Ribbons in Coronal Null-Point
  Topology}}.
\newblock \emph{\bibinfo{journal}{\apj}} \textbf{\bibinfo{volume}{700}},
  \bibinfo{pages}{559--578} (\bibinfo{year}{2009}).

\bibitem{Wang&Liu2012}
\bibinfo{author}{{Wang}, H.} \& \bibinfo{author}{{Liu}, C.}
\newblock \bibinfo{title}{{Circular Ribbon Flares and Homologous Jets}}.
\newblock \emph{\bibinfo{journal}{\apj}} \textbf{\bibinfo{volume}{760}},
  \bibinfo{pages}{101} (\bibinfo{year}{2012}).

\bibitem{Sun2013}
\bibinfo{author}{{Sun}, X.} \emph{et~al.}
\newblock \bibinfo{title}{{Hot Spine Loops and the Nature of a Late-phase Solar
  Flare}}.
\newblock \emph{\bibinfo{journal}{\apj}} \textbf{\bibinfo{volume}{778}},
  \bibinfo{pages}{139} (\bibinfo{year}{2013}).

\bibitem{Jiang2014}
\bibinfo{author}{{Jiang}, C.}, \bibinfo{author}{{Wu}, S.~T.},
  \bibinfo{author}{{Feng}, X.} \& \bibinfo{author}{{Hu}, Q.}
\newblock \bibinfo{title}{{Formation and Eruption of an Active Region Sigmoid.
  I. A Study by Nonlinear Force-free Field Modeling}}.
\newblock \emph{\bibinfo{journal}{\apj}} \textbf{\bibinfo{volume}{780}},
  \bibinfo{pages}{55} (\bibinfo{year}{2014}).

\bibitem{Vemareddy&Wiegelmann2014}
\bibinfo{author}{{Vemareddy}, P.} \& \bibinfo{author}{{Wiegelmann}, T.}
\newblock \bibinfo{title}{{Quasi-static Three-dimensional Magnetic Field
  Evolution in Solar Active Region NOAA 11166 Associated with an X1.5 Flare}}.
\newblock \emph{\bibinfo{journal}{\apj}} \textbf{\bibinfo{volume}{792}},
  \bibinfo{pages}{40} (\bibinfo{year}{2014}).

\bibitem{Joshi2015}
\bibinfo{author}{{Joshi}, N.~C.} \emph{et~al.}
\newblock \bibinfo{title}{{The Role of Erupting Sigmoid in Triggering a Flare
  with Parallel and Large-scale Quasi-circular Ribbons}}.
\newblock \emph{\bibinfo{journal}{\apj}} \textbf{\bibinfo{volume}{812}},
  \bibinfo{pages}{50} (\bibinfo{year}{2015}).

\bibitem{Liu2015}
\bibinfo{author}{{Liu}, C.} \emph{et~al.}
\newblock \bibinfo{title}{{A Circular-ribbon Solar Flare Following an
  Asymmetric Filament Eruption}}.
\newblock \emph{\bibinfo{journal}{\apjl}} \textbf{\bibinfo{volume}{812}},
  \bibinfo{pages}{L19} (\bibinfo{year}{2015}).

\bibitem{hoeksema14}
\bibinfo{author}{{Hoeksema}, J.~T.} \emph{et~al.}
\newblock \bibinfo{title}{{The Helioseismic and Magnetic Imager (HMI) Vector
  Magnetic Field Pipeline: Overview and Performance}}.
\newblock \emph{\bibinfo{journal}{\solphys}} \textbf{\bibinfo{volume}{289}},
  \bibinfo{pages}{3483--3530} (\bibinfo{year}{2014}).

\bibitem{pesnell12}
\bibinfo{author}{{Pesnell}, W.~D.}, \bibinfo{author}{{Thompson}, B.~J.} \&
  \bibinfo{author}{{Chamberlin}, P.~C.}
\newblock \bibinfo{title}{{The Solar Dynamics Observatory (SDO)}}.
\newblock \emph{\bibinfo{journal}{\solphys}} \textbf{\bibinfo{volume}{275}},
  \bibinfo{pages}{3--15} (\bibinfo{year}{2012}).

\bibitem{Bobra2014}
\bibinfo{author}{{Bobra}, M.~G.} \emph{et~al.}
\newblock \bibinfo{title}{{The Helioseismic and Magnetic Imager (HMI) Vector
  Magnetic Field Pipeline: SHARPs - Space-Weather HMI Active Region Patches}}.
\newblock \emph{\bibinfo{journal}{\solphys}} \textbf{\bibinfo{volume}{289}},
  \bibinfo{pages}{3549--3578} (\bibinfo{year}{2014}).

\bibitem{lemen12}
\bibinfo{author}{{Lemen}, J.~R.}, \bibinfo{author}{{Title}, A.~M.},
  \bibinfo{author}{{Akin}, D.~J.} \emph{et~al.}
\newblock \bibinfo{title}{{The Atmospheric Imaging Assembly (AIA) on the Solar
  Dynamics Observatory (SDO)}}.
\newblock \emph{\bibinfo{journal}{\solphys}} \textbf{\bibinfo{volume}{275}},
  \bibinfo{pages}{17--40} (\bibinfo{year}{2012}).

\bibitem{ODwyer2010}
\bibinfo{author}{{O'Dwyer}, B.}, \bibinfo{author}{{Del Zanna}, G.},
  \bibinfo{author}{{Mason}, H.~E.}, \bibinfo{author}{{Weber}, M.~A.} \&
  \bibinfo{author}{{Tripathi}, D.}
\newblock \bibinfo{title}{{SDO/AIA response to coronal hole, quiet Sun, active
  region, and flare plasma}}.
\newblock \emph{\bibinfo{journal}{\aap}} \textbf{\bibinfo{volume}{521}},
  \bibinfo{pages}{A21} (\bibinfo{year}{2010}).

\bibitem{lin02}
\bibinfo{author}{{Lin}, R.~P.}, \bibinfo{author}{{Dennis}, B.~R.},
  \bibinfo{author}{{Hurford}, G.~J.} \emph{et~al.}
\newblock \bibinfo{title}{{The Reuven Ramaty High-Energy Solar Spectroscopic
  Imager (RHESSI)}}.
\newblock \emph{\bibinfo{journal}{\solphys}} \textbf{\bibinfo{volume}{210}},
  \bibinfo{pages}{3--32} (\bibinfo{year}{2002}).

\bibitem{Machado&Henoux1982}
\bibinfo{author}{{Machado}, M.~E.} \& \bibinfo{author}{{Henoux}, J.-C.}
\newblock \bibinfo{title}{{Chromospheric effects of XUV radiation emitted
  during solar flares}}.
\newblock \emph{\bibinfo{journal}{\aap}} \textbf{\bibinfo{volume}{108}},
  \bibinfo{pages}{61--68} (\bibinfo{year}{1982}).

\bibitem{Doyle&Phillips1992}
\bibinfo{author}{{Doyle}, J.~G.} \& \bibinfo{author}{{Phillips}, K.~J.~H.}
\newblock \bibinfo{title}{{Excitation of the solar flare far-ultraviolet
  continuum by line irradiation}}.
\newblock \emph{\bibinfo{journal}{\aap}} \textbf{\bibinfo{volume}{257}},
  \bibinfo{pages}{773--776} (\bibinfo{year}{1992}).

\bibitem{Yang2015}
\bibinfo{author}{{Yang}, S.}, \bibinfo{author}{{Zhang}, J.} \&
  \bibinfo{author}{{Xiang}, Y.}
\newblock \bibinfo{title}{{Magnetic Reconnection between Small-scale Loops
  Observed with the New Vacuum Solar Telescope}}.
\newblock \emph{\bibinfo{journal}{\apjl}} \textbf{\bibinfo{volume}{798}},
  \bibinfo{pages}{L11} (\bibinfo{year}{2015}).

\bibitem{Fletcher2013}
\bibinfo{author}{{Fletcher}, L.}, \bibinfo{author}{{Hannah}, I.~G.},
  \bibinfo{author}{{Hudson}, H.~S.} \& \bibinfo{author}{{Innes}, D.~E.}
\newblock \bibinfo{title}{{Flare Ribbon Energetics in the Early Phase of an SDO
  Flare}}.
\newblock \emph{\bibinfo{journal}{\apj}} \textbf{\bibinfo{volume}{771}},
  \bibinfo{pages}{104} (\bibinfo{year}{2013}).

\bibitem{Simoes2015}
\bibinfo{author}{{Sim{\~o}es}, P.~J.~A.}, \bibinfo{author}{{Graham}, D.~R.} \&
  \bibinfo{author}{{Fletcher}, L.}
\newblock \bibinfo{title}{{Impulsive Heating of Solar Flare Ribbons Above 10
  MK}}.
\newblock \emph{\bibinfo{journal}{\solphys}} \textbf{\bibinfo{volume}{290}},
  \bibinfo{pages}{3573--3591} (\bibinfo{year}{2015}).

\bibitem{Polito2016}
\bibinfo{author}{{Polito}, V.} \emph{et~al.}
\newblock \bibinfo{title}{{Simultaneous IRIS and Hinode/EIS Observations and
  Modelling of the 2014 October 27 X2.0 Class Flare}}.
\newblock \emph{\bibinfo{journal}{\apj}} \textbf{\bibinfo{volume}{816}},
  \bibinfo{pages}{89} (\bibinfo{year}{2016}).

\bibitem{Aschwanden1999}
\bibinfo{author}{{Aschwanden}, M.~J.}, \bibinfo{author}{{Kosugi}, T.},
  \bibinfo{author}{{Hanaoka}, Y.}, \bibinfo{author}{{Nishio}, M.} \&
  \bibinfo{author}{{Melrose}, D.~B.}
\newblock \bibinfo{title}{{Quadrupolar Magnetic Reconnection in Solar Flares.
  I. Three-dimensional Geometry Inferred from Yohkoh Observations}}.
\newblock \emph{\bibinfo{journal}{\apj}} \textbf{\bibinfo{volume}{526}},
  \bibinfo{pages}{1026--1045} (\bibinfo{year}{1999}).

\bibitem{Solanki2003}
\bibinfo{author}{{Solanki}, S.~K.}
\newblock \bibinfo{title}{{Sunspots: An overview}}.
\newblock \emph{\bibinfo{journal}{\aapr}} \textbf{\bibinfo{volume}{11}},
  \bibinfo{pages}{153--286} (\bibinfo{year}{2003}).

\bibitem{Zwaan1987}
\bibinfo{author}{{Zwaan}, C.}
\newblock \bibinfo{title}{{Elements and patterns in the solar magnetic field}}.
\newblock \emph{\bibinfo{journal}{\araa}} \textbf{\bibinfo{volume}{25}},
  \bibinfo{pages}{83--111} (\bibinfo{year}{1987}).

\bibitem{Liu2014}
\bibinfo{author}{{Liu}, R.} \emph{et~al.}
\newblock \bibinfo{title}{{An Unorthodox X-Class Long-duration Confined
  Flare}}.
\newblock \emph{\bibinfo{journal}{\apj}} \textbf{\bibinfo{volume}{790}},
  \bibinfo{pages}{8} (\bibinfo{year}{2014}).

\bibitem{Parnell1996}
\bibinfo{author}{{Parnell}, C.~E.}, \bibinfo{author}{{Smith}, J.~M.},
  \bibinfo{author}{{Neukirch}, T.} \& \bibinfo{author}{{Priest}, E.~R.}
\newblock \bibinfo{title}{{The structure of three-dimensional magnetic neutral
  points}}.
\newblock \emph{\bibinfo{journal}{Physics of Plasmas}}
  \textbf{\bibinfo{volume}{3}}, \bibinfo{pages}{759--770}
  (\bibinfo{year}{1996}).

\bibitem{Haynes&Parnell2007}
\bibinfo{author}{{Haynes}, A.~L.} \& \bibinfo{author}{{Parnell}, C.~E.}
\newblock \bibinfo{title}{{A trilinear method for finding null points in a
  three-dimensional vector space}}.
\newblock \emph{\bibinfo{journal}{Physics of Plasmas}}
  \textbf{\bibinfo{volume}{14}}, \bibinfo{pages}{082107--082107}
  (\bibinfo{year}{2007}).

\bibitem{Priest2014book}
\bibinfo{author}{{Priest}, E.}
\newblock \emph{\bibinfo{title}{{Magnetohydrodynamics of the Sun}}}
  (\bibinfo{publisher}{Cambridge University Press}, \bibinfo{year}{2014}),
  \bibinfo{edition}{1st} edn.

\bibitem{wiegelmann04}
\bibinfo{author}{{Wiegelmann}, T.}
\newblock \bibinfo{title}{{Optimization code with weighting function for the
  reconstruction of coronal magnetic fields}}.
\newblock \emph{\bibinfo{journal}{\solphys}} \textbf{\bibinfo{volume}{219}},
  \bibinfo{pages}{87--108} (\bibinfo{year}{2004}).

\bibitem{wiegelmann06}
\bibinfo{author}{{Wiegelmann}, T.}, \bibinfo{author}{{Inhester}, B.} \&
  \bibinfo{author}{{Sakurai}, T.}
\newblock \bibinfo{title}{{Preprocessing of Vector Magnetograph Data for a
  Nonlinear Force-Free Magnetic Field Reconstruction}}.
\newblock \emph{\bibinfo{journal}{\solphys}} \textbf{\bibinfo{volume}{233}},
  \bibinfo{pages}{215--232} (\bibinfo{year}{2006}).

\bibitem{schuck08}
\bibinfo{author}{{Schuck}, P.~W.}
\newblock \bibinfo{title}{{Tracking Vector Magnetograms with the Magnetic
  Induction Equation}}.
\newblock \emph{\bibinfo{journal}{\apj}} \textbf{\bibinfo{volume}{683}},
  \bibinfo{pages}{1134--1152} (\bibinfo{year}{2008}).

\bibitem{liuy14}
\bibinfo{author}{{Liu}, Y.} \emph{et~al.}
\newblock \bibinfo{title}{{Magnetic Helicity in Emerging Solar Active
  Regions}}.
\newblock \emph{\bibinfo{journal}{\apj}} \textbf{\bibinfo{volume}{785}},
  \bibinfo{pages}{13} (\bibinfo{year}{2014}).

\bibitem{liuy12}
\bibinfo{author}{{Liu}, Y.} \& \bibinfo{author}{{Schuck}, P.~W.}
\newblock \bibinfo{title}{{Magnetic Energy and Helicity in Two Emerging Active
  Regions in the Sun}}.
\newblock \emph{\bibinfo{journal}{\apj}} \textbf{\bibinfo{volume}{761}},
  \bibinfo{pages}{105} (\bibinfo{year}{2012}).

\bibitem{kusano02}
\bibinfo{author}{{Kusano}, K.}, \bibinfo{author}{{Maeshiro}, T.},
  \bibinfo{author}{{Yokoyama}, T.} \& \bibinfo{author}{{Sakurai}, T.}
\newblock \bibinfo{title}{{Measurement of Magnetic Helicity Injection and Free
  Energy Loading into the Solar Corona}}.
\newblock \emph{\bibinfo{journal}{\apj}} \textbf{\bibinfo{volume}{577}},
  \bibinfo{pages}{501--512} (\bibinfo{year}{2002}).

\bibitem{Liu2016}
\bibinfo{author}{{Liu}, R.} \emph{et~al.}
\newblock \bibinfo{title}{{Structure, Stability, and Evolution of Magnetic Flux
  Ropes from the Perspective of Magnetic Twist}}.
\newblock \emph{\bibinfo{journal}{\apj}} \textbf{\bibinfo{volume}{818}},
  \bibinfo{pages}{148} (\bibinfo{year}{2016}).

\bibitem{Alissandrakis1981}
\bibinfo{author}{{Alissandrakis}, C.~E.}
\newblock \bibinfo{title}{{On the computation of constant alpha force-free
  magnetic field}}.
\newblock \emph{\bibinfo{journal}{\aap}} \textbf{\bibinfo{volume}{100}},
  \bibinfo{pages}{197--200} (\bibinfo{year}{1981}).

\end{thebibliography}

\begin{addendum}
 \item [Acknowledgments] We thank the SDO team for the vector magnetic field and EUV imaging data, P.W. Schuck for the flow tracking code, T. Wiegelmann for the NLFFF code, X. Cheng for helpful comments.  R.L. acknowledges the support from the Thousand Young Talents Program of China and NSFC 41474151. Y.W. acknowledges the support from NSFC 41131065 and 41574165. This work was also supported by NSFC 41421063, CAS Key Research Program KZZD-EW-01-4, and the fundamental research funds for the central universities.
 \item [Correspondence] Correspondence and requests for materials should be addressed to R.L.~(rliu@ustc.edu.cn).
\end{addendum}

\clearpage
\begin{figure} 
	\centering
	\includegraphics[width=\hsize]{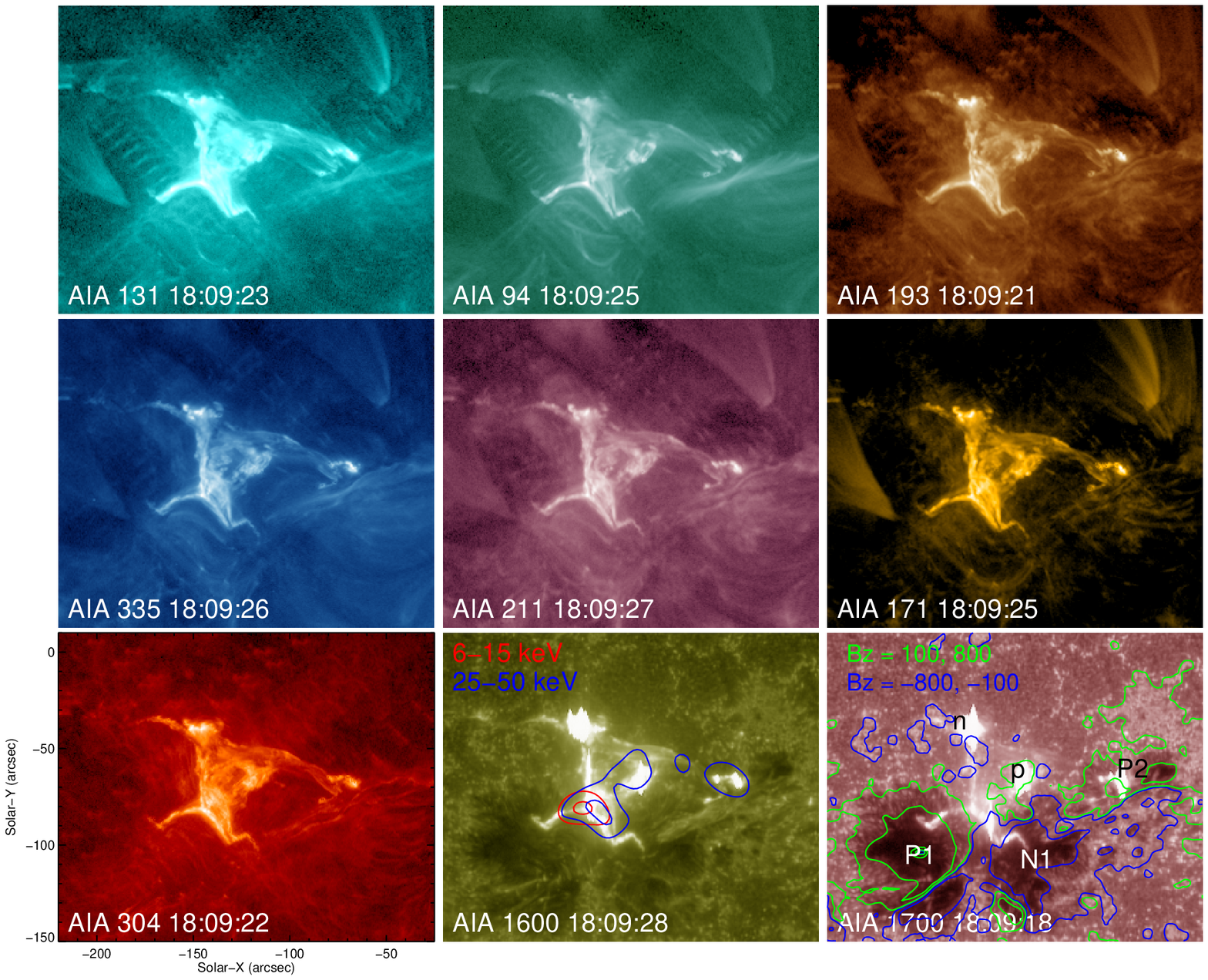}
	\caption{\small Snapshots of the X-shaped M3.1 flare at 18:11 UT on 2014 February 2 in AIA's nine UV/EUV passbands. Heliocentric-cartesian coordinates are adopted so that $X$ ($Y$) increases towards the solar west (north). The AIA 1600~{\AA} image is superimposed by RHESSI HXR contours at 6--12 (red) and 25--50 (blue) keV, with the contour levels set at 50 and 90 percent of the maximum intensity. The AIA 1700~{\AA} image is superimposed by the contours of the line-of-sight component of the photospheric magnetic field. The green and blue colors indicate positive and negative polarities, respectively, with the contour levels set at $\pm100$ and $\pm800$ Gauss. An animation of AIA images is provided in \suppmv~1. \label{fig:flare3}}
\end{figure}

\begin{figure}
	\centering
	\includegraphics[width=\hsize]{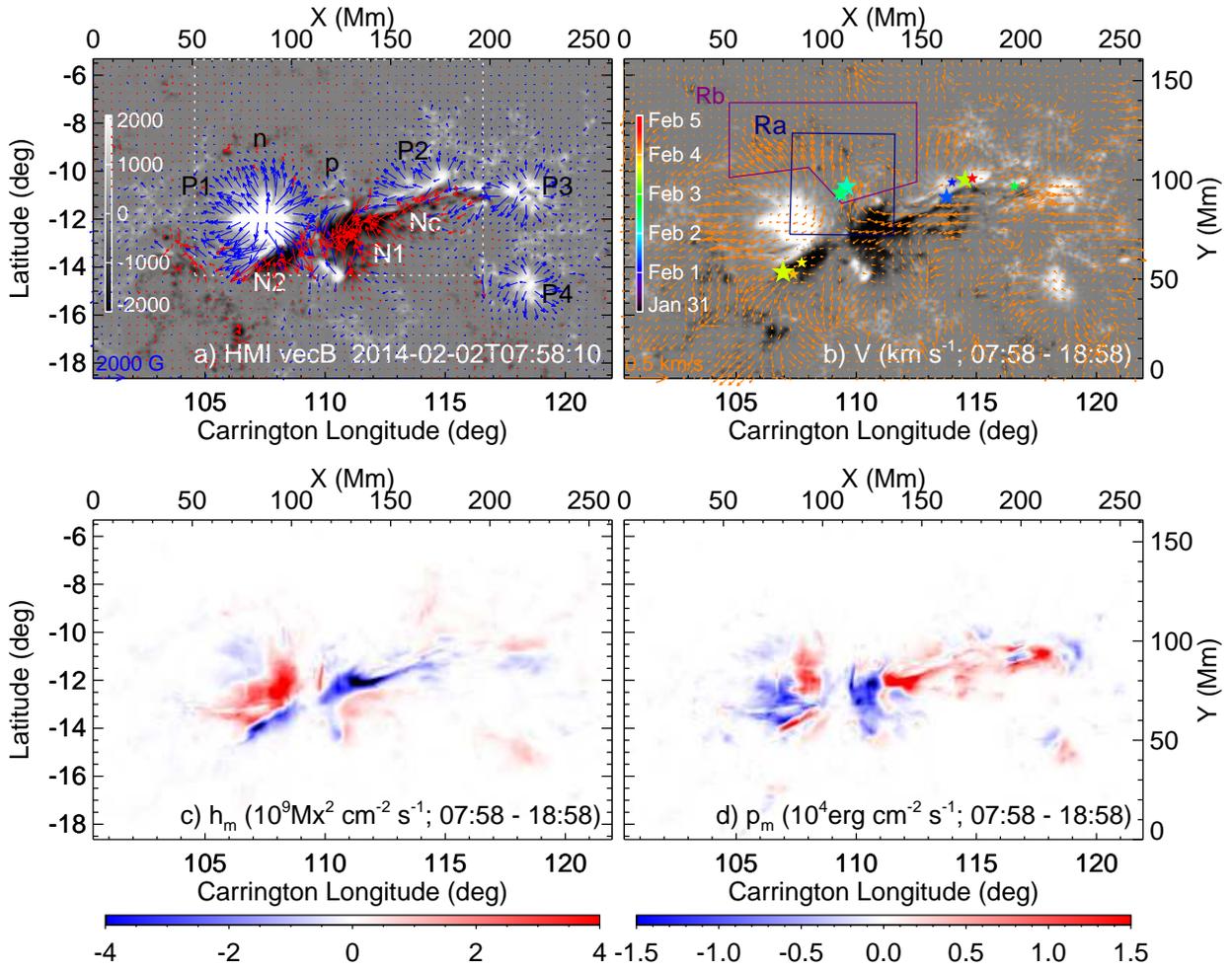}
	\caption{\small HMI observation of AR 11967. (a) Vector magnetogram of AR 11967 immediately before the occurrence of the first XMF at 08:20 UT on 2014 February 2. White and black colors refer to positive and negative $B_z$, respectively, which are scaled to $\pm2000$ G (see the color bar). Red (blue) arrows represent the tangential field components that originate from negative (positive) $B_z$ elements. (b--d) Maps of photospheric flows, helicity and Poynting flux density averaged over a time period of 11 hrs covering the three XMFs. In Panel (b) the locations of all the major flares from January 31 to February 5, as given in Table~\ref{table}, are marked with star symbols. The colors and sizes of the symbols indicate different times (see the color bar) and flare magnitudes, respectively. The largest size corresponds to the M5.2 flare at 03:57 UT on 2014 February 4. The evolution of the vector magnetic field, flow field, helicity and Poynting flux density is provided in \suppmv~2. \label{fig:bfield}}
\end{figure}

\begin{figure}
  \centering
  \includegraphics[width=0.9\hsize]{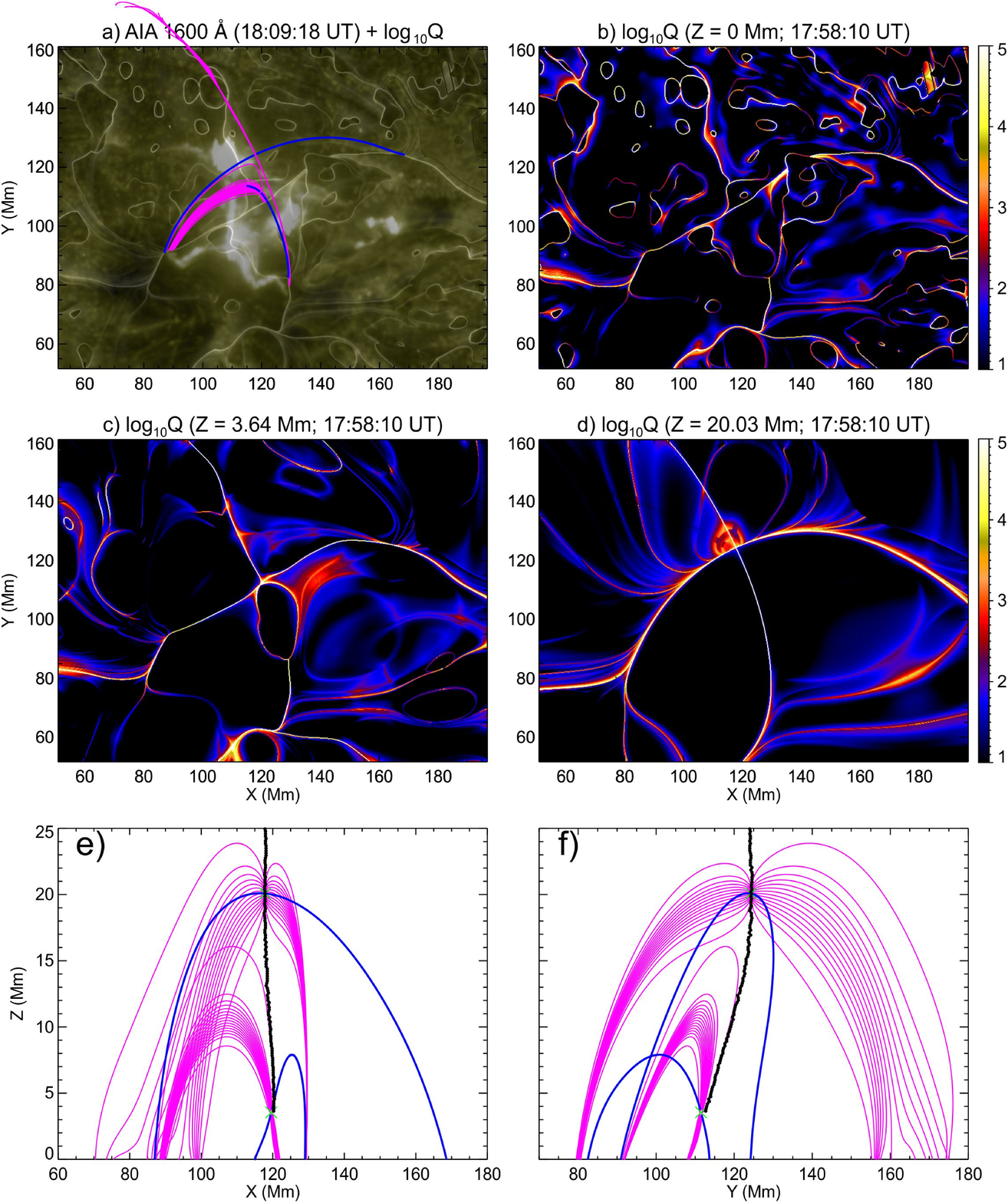}
  \caption{\small Magnetic topology prior to the XMF at 18:11 UT on 2014 February 2. a) A UV 1600~{\AA} image taken during the impulsive phase of the flare (same as in Figure~\ref{fig:flare3}) and remapped with the CEA projection is blended with the $\log Q$ map on the photosphere in Panel (b),  which is calculated with the HMI magnetogram at 17:58 UT. Panels (c) and (d) show two cuts of $\log Q$ at different heights just above the lower null and below the upper null, respectively. Field lines traced in the neighborhood of the double null, with fan (spine) field lines in magenta (blue), are projected in X-Y (Panel a), X-Z (Panel e), and Y-Z (Panel f) planes. The black dots in (e) and (f) indicate the visually determined center of the X shape in $\log Q$ at different heights. The X shape is complicated below the lower null, hence its central position is not shown here. An animation of $\log Q$ at different heights is provided in \suppmv~3.  \label{fig:null}}
\end{figure}

\begin{figure}
  \centering
  \includegraphics[width=\hsize]{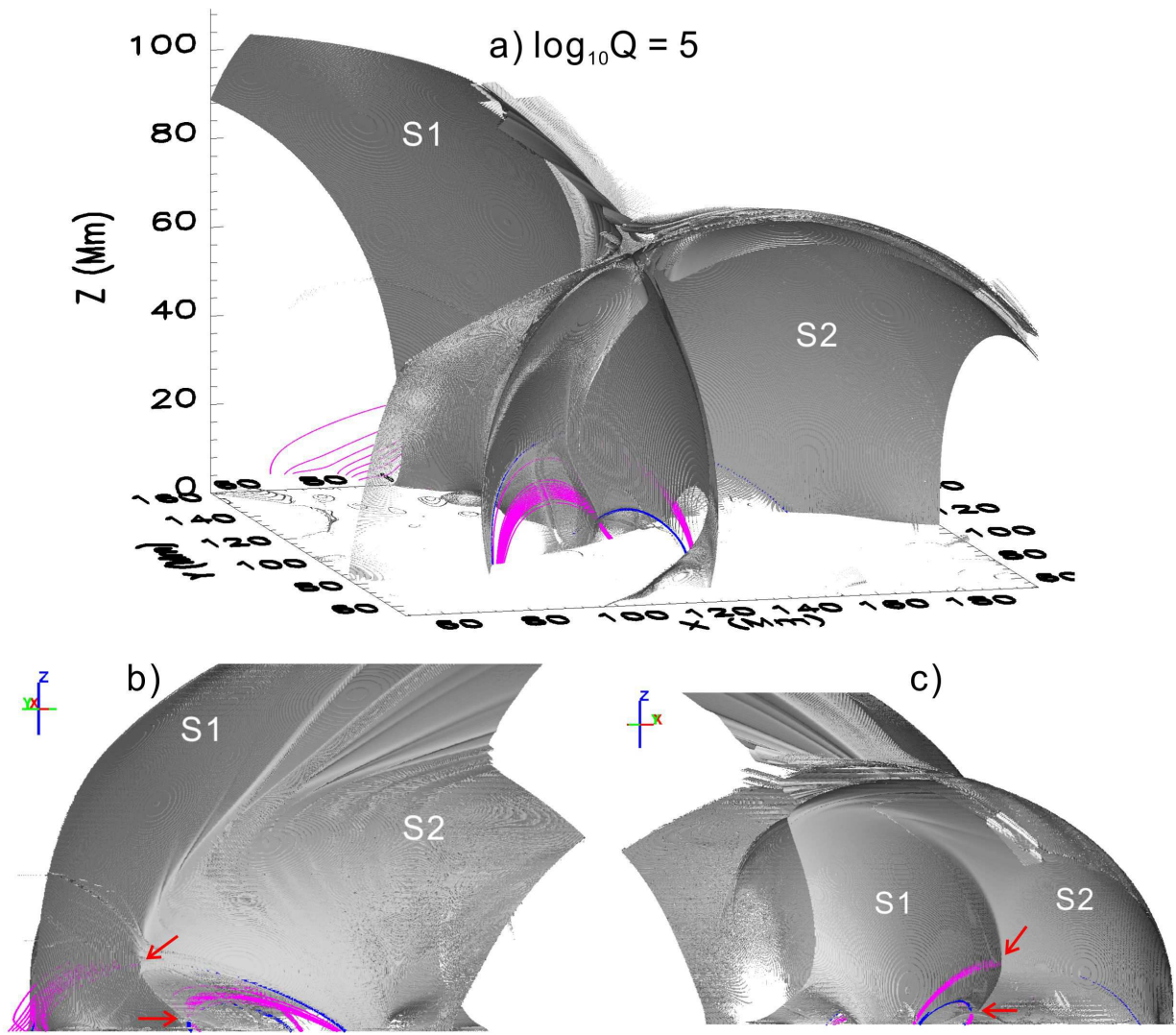}
  \caption{\small QSLs prior to the XMF at 18:11 UT on 2014 February 2. a) Isosurfaces of $\log Q =5$, superimposed by field lines traced in the neighbourhood of the double null, with fan (spine) field lines in magenta (blue).  Bottom panels show two side views of the isosurfaces from the east (b) and the southwest (c), respectively. The null locations are roughly indicated by red arrows. A 360 deg side view of the isosurfaces is provided in \suppmv~4.  \label{fig:qsl}}
\end{figure}

\begin{figure}
  \centering
  \includegraphics[width=\hsize]{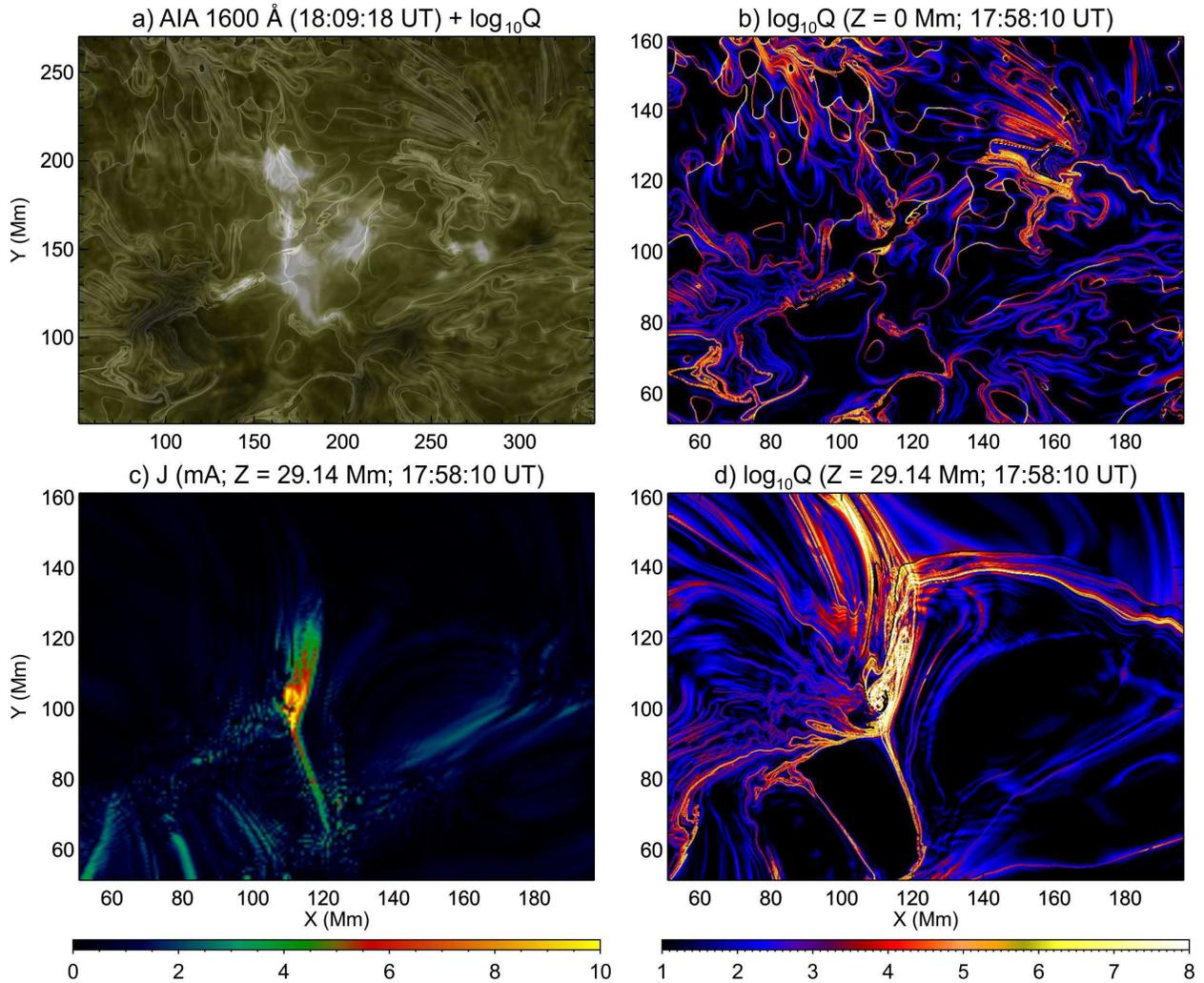}
  \caption{\small Magnetic topology as revealed by a NLFFF model for the XMF at 18:11 UT on 2014 February 2. Panel (a) shows a blend of UV 1600~{\AA} image and the photospheric $\log Q$ map, the latter of which is shown in Panel (b). Panels (c) and (d) show a horizontal cut of the electric current and $\log Q$ at $Z=29$ Mm, respectively. An animation of $\log Q$ at different heights is provided in \suppmv~5. \label{fig:nlff}}
\end{figure}

\begin{table}
	\centering
	\begin{threeparttable} 
		\caption{List of Major Flares (M-class and above) in AR 11967}
		\begin{tabular}{cccccc}
			\toprule
			Date & Class & Peak & Carr\_Lon & Carr\_Lat & Instrument \\
			\midrule
			2014/02/01 & M1.0 & 01:25 & 113.987 & -10.463 & AIA \\
			2014/02/01 & M3.0 & 07:23 & 113.759 & -11.086 & RHESSI \\
			\textbf{2014/02/02} & \textbf{M2.2} & \textbf{08:20} & \textbf{109.400} & \textbf{-10.995} & \textbf{RHESSI}  \\
			\textbf{2014/02/02} & \textbf{M4.4} & \textbf{09:31} & \textbf{109.602} & \textbf{-10.641} & \textbf{AIA } \\
			\textbf{2014/02/02} & \textbf{M3.1} & \textbf{18:11} & \textbf{109.325} & \textbf{-10.898} & \textbf{RHESSI} \\
			2014/02/02 & M1.3 & 22:04 & 116.584 & -10.625 & RHESSI \\
			2014/02/04 & M3.8 & 01:23 & 114.536 & -10.355 & RHESSI \\
			2014/02/04 & M5.2 & 04:00 & 106.954 & -14.215 & AIA \\
			2014/02/04 & M1.4 & 09:49 & 107.728 & -13.806 & RHESSI \\
			2014/02/04 & M1.5 & 16:02 & 107.377 & -14.291	& RHESSI \\
			2014/02/05 & M1.3 & 16:20 & 114.830 & -10.301 & AIA \\
			\bottomrule
		\end{tabular}
		\begin{tablenotes}
			\small
			\item The Carrington longitudes and latitudes of the flare locations are given by the peak emission loci of RHESSI HXR sources at 6--12 keV or by the peak 131~{\AA} emission loci in the flaring regions when RHESSI data were not available. The three X-shaped flares are shown in boldface. All the events were `confined' without being associated with a coronal mass ejection (CME). \label{table}
		\end{tablenotes}
	\end{threeparttable}
\end{table}

\setcounter{figure}{0}
\renewcommand{\figurename}{\textbf{Supp.\,Fig.}}
\renewcommand{\tablename}{\textbf{Supp.\,Table}}

\clearpage
\section*{Supplementary Notes} \label{append:observation}
\begin{figure}[!htbp] 
	\centering
	\includegraphics[width=\hsize]{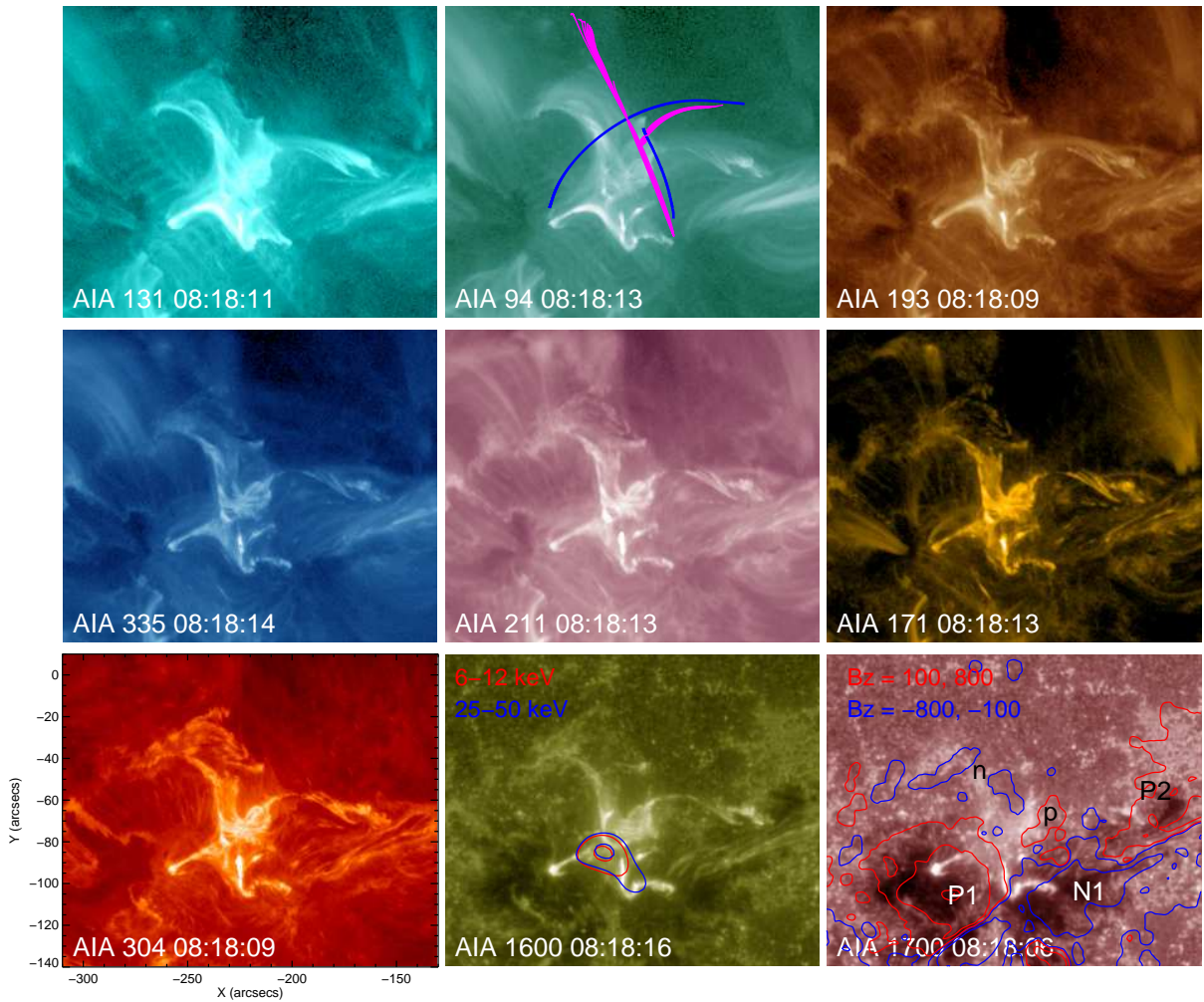}
	\caption{\small A snapshot of the M2.2 flare at 08:20 UT on 2014 Feburary 2 in AIA's nine UV/EUV passbands. The spine (blue) and fan (magenta) field lines of two coronal null points are projected onto the AIA 94~{\AA} image. The AIA 1600~{\AA} image is superimposed by RHESSI HXR contours at 6--12 (red) and 25--50 (blue) keV, with the contour levels set at 50 and 90 percent of the maximum intensity. The AIA 1700~{\AA} image is superimposed by the contours of the LOS component of the photospheric magnetic field. The red and blue colors indicate positive and negative polarities, respectively, with the contour levels set at $\pm100$ and $\pm800$ Gauss. An animation of AIA images is provided in \suppmv~6. \label{suppfig:flare1}}
\end{figure}

\begin{figure}[!htbp]
	\centering
	\includegraphics[width=\hsize]{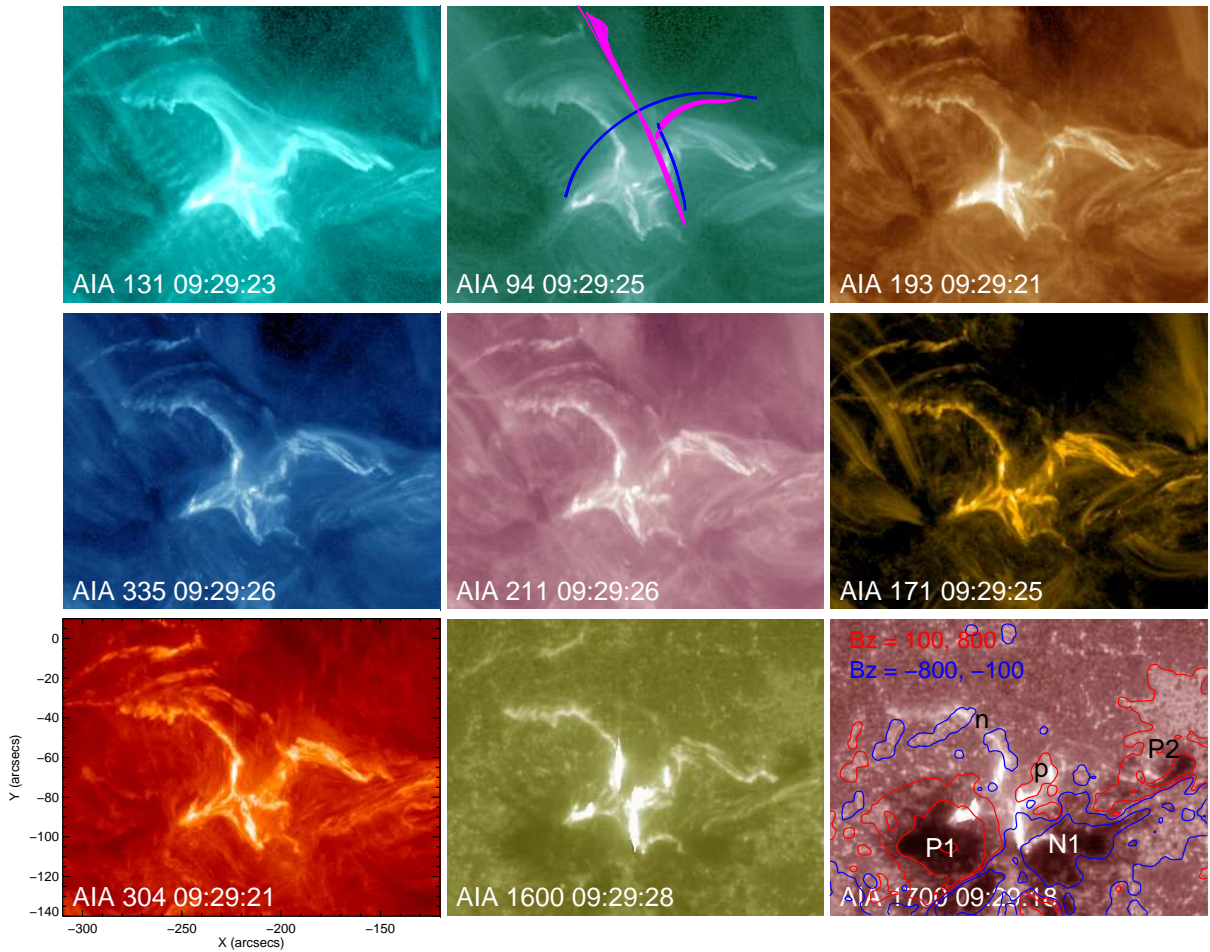}
	\caption{\small A snapshot of the M4.4 flare at 09:31 UT on 2014 Feburary 2 in AIA's nine UV/EUV passbands. The spine (blue) and fan (magenta) field lines of two coronal null points are projected onto the AIA 94~{\AA} image. The AIA 1700~{\AA} image is superimposed by the contours of the LOS component of the photospheric magnetic field. The red and blue colors indicate positive and negative polarities, respectively, with the contour levels set at $\pm100$ and $\pm800$ Gauss. An animation of AIA images is provided in \suppmv~7. \label{suppfig:flare2}}
\end{figure}

\begin{figure}[!htbp]
	\centering
    \includegraphics[width=\hsize]{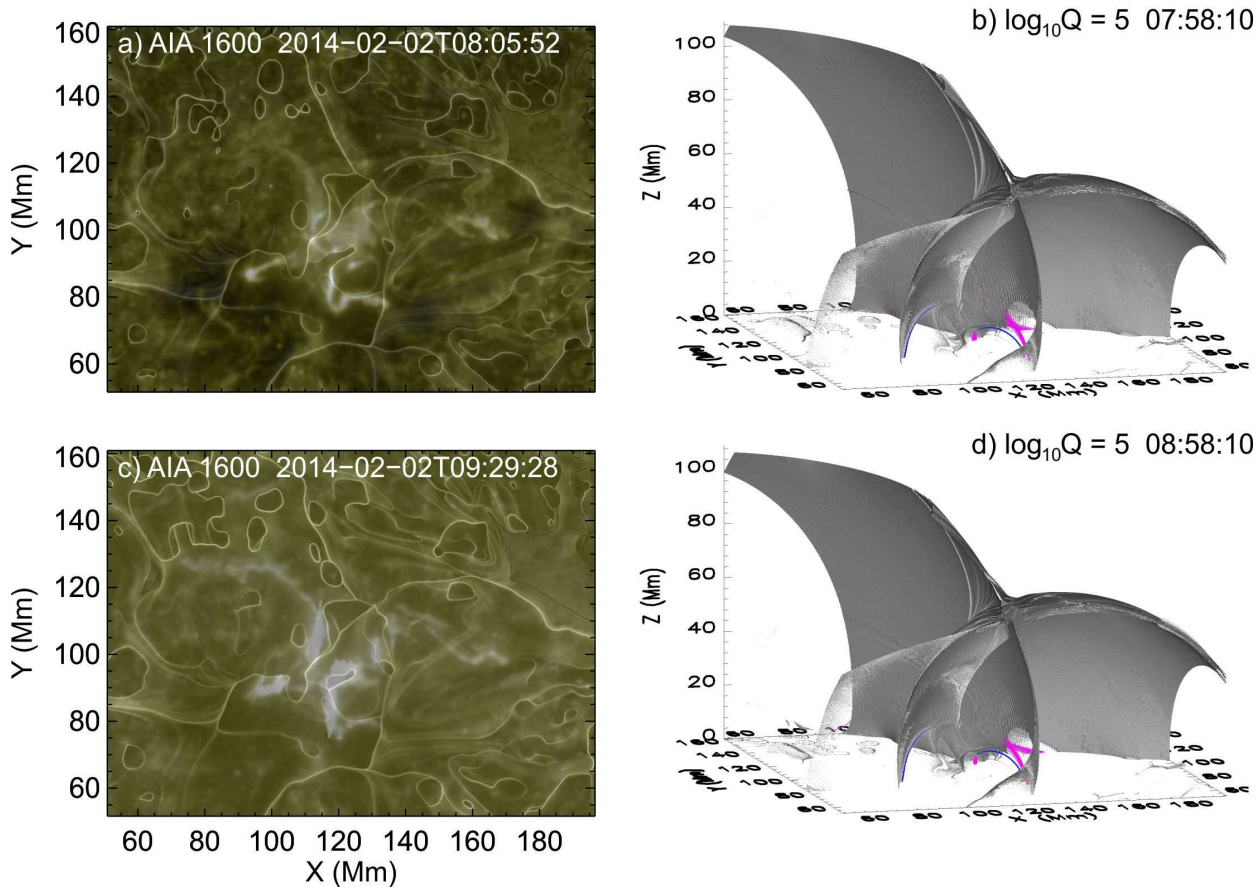}
	\caption{\small Magnetic topology relevant to the M2.2 flare at 08:20 UT and M4.4 flare at 09:31 UT on 2014 February 2. Left column: A UV 1600~{\AA} image taken during the impulsive phase of the flare and remapped with the CEA projection is blended with the $\log Q$ map calculated for the HMI magnetogram acquired at approximately the same time. Right column: Isosurfaces of $\log Q =5$, overplotted by field lines traced in the neighborhood of the double null, with `fan' lines in magenta and `spine' lines in blue. A 360 deg side view of the $\log Q=5$ isosurfaces in Panels (b) and (d) is provided in \suppmv s~8 and 9, respectively.   \label{suppfig:qsl}}  
\end{figure}

\suppfig s~\ref{suppfig:flare1} (\suppmv~6) and \ref{suppfig:flare2} (\suppmv~7) show the AIA observations of the other two XMFs occurring at 08:20 UT and 09:31 UT on 2014 February 2, respectively. The magnetic topology relevant to the X-shaped ribbons in these two flares was similar to the flare at 18:11 UT (Figures~\ref{fig:null}--\ref{fig:nlff}), with a double null embedded in the intersection of two high-$Q$ surfaces (\suppfig~\ref{suppfig:qsl} and \suppmv s~8 and 9). 

\begin{figure}[!htbp]
	\centering
	\includegraphics[width=\hsize]{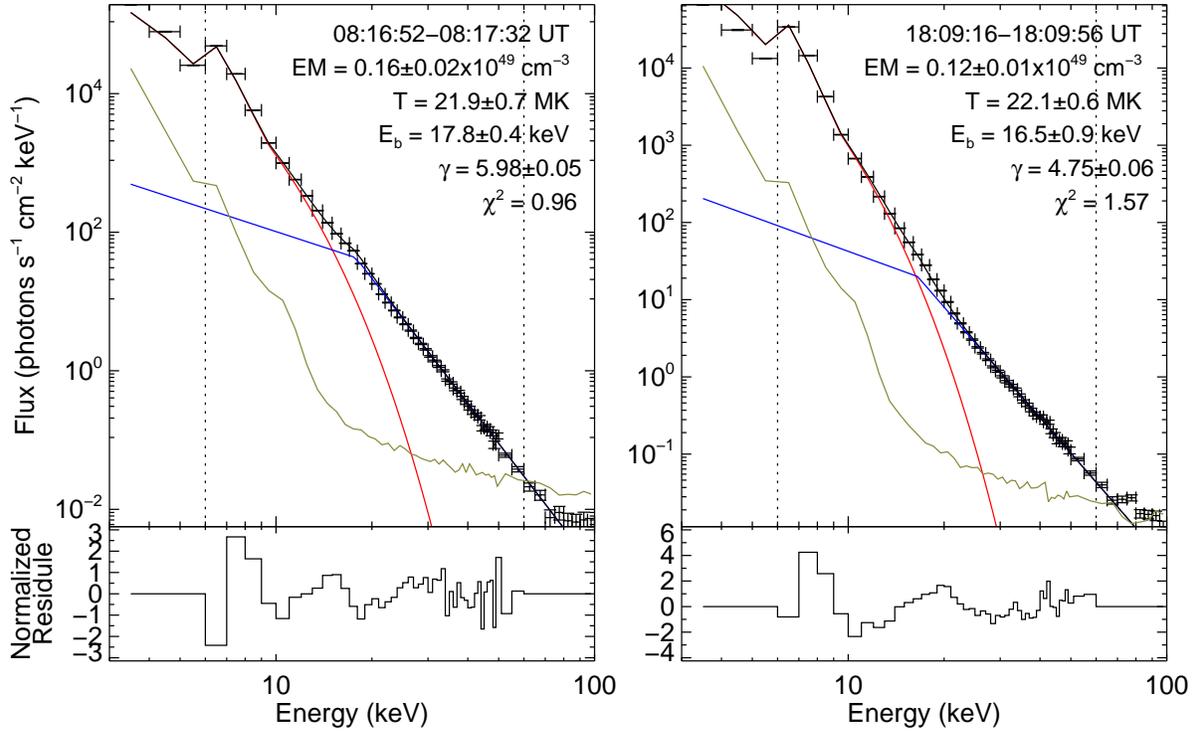}
	\caption{\small Flare spectroscopy for the M2.2 flare at 08:20 UT and M3.1 flare at 18:11 UT on 2014 February 2. RHESSI spectra at the flare peak are fitted in an energy range of 6--100 keV (denoted by dotted lines) with an exponential thermal function (red) and a broken power-law function (blue) with the spectral index $\gamma$ below the broken energy $E_b$ being fixed at 1.5. The background is shown in gray. \label{suppfig:spectra}}
\end{figure}

The XMFs at 08:20 UT (\suppfig~\ref{suppfig:flare1}) and 18:11 UT (Figure~\ref{fig:flare3}) were observed in HXRs by RHESSI. In both flares, we integrated the HXR spectra over 40 s around the peak of the nonthermal HXR bursts (\suppfig~\ref{suppfig:spectra}) and found that they can be well fitted with an exponential thermal function (red) and a nonthermal power-law function (blue). 

\clearpage
\section*{Supplementary Figures}

\begin{figure}[!htbp]
	\centering
	\includegraphics[width=\hsize]{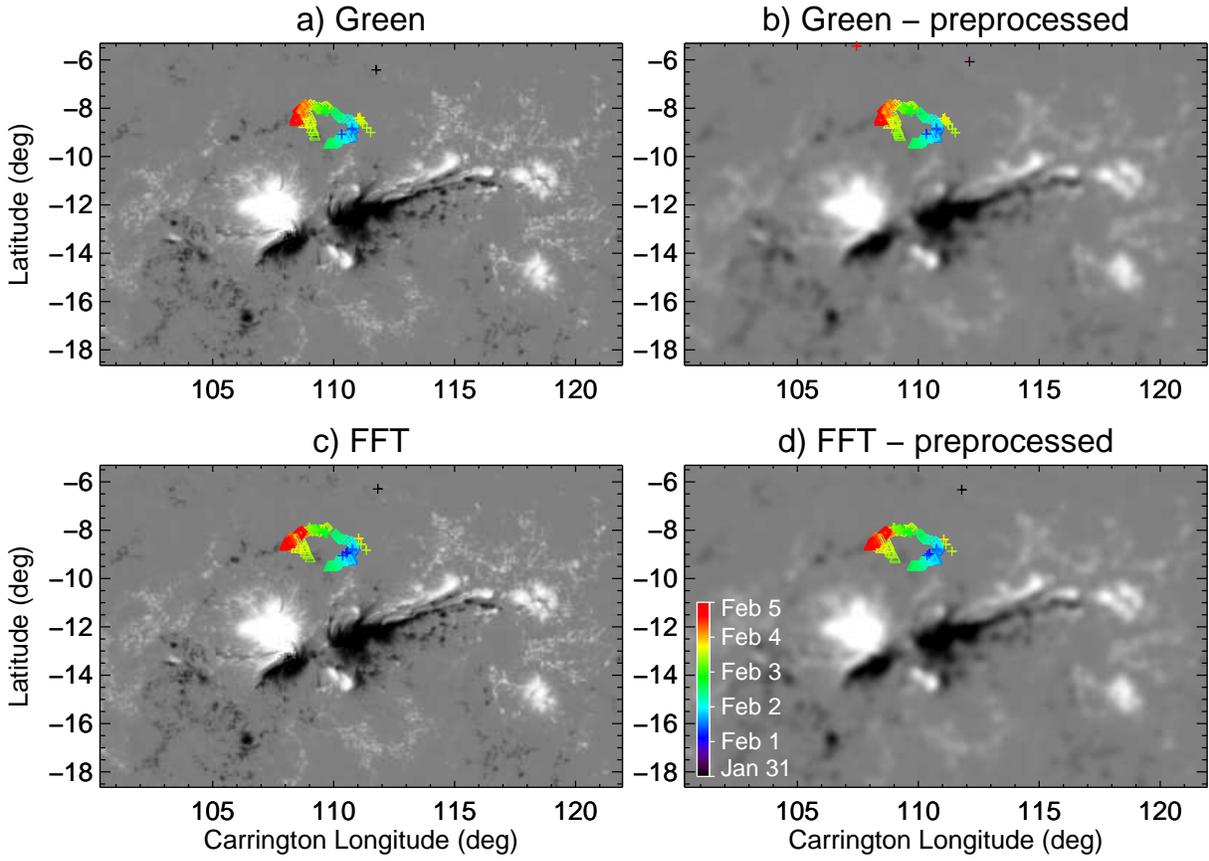}
	\caption{\small Null locations projected on the photospheric boundary used for the potential field extrapolation. The nulls are identified in potential fields constructed with different methods (Green function vs. fast Fourier transform) and different boundary conditions (original vs. pre-processsed). Same as in Figure~\ref{fig:ltc}, the lower (upper) null is marked with a triangle (diamond) for double nulls, while a single null is marked by a `+' symbol. The color of symbols indicates the time when the null was present (see the color bar). \label{suppfig:nullxy}}
\end{figure}

\begin{figure}[!htbp]
	\centering
	\includegraphics[width=\hsize]{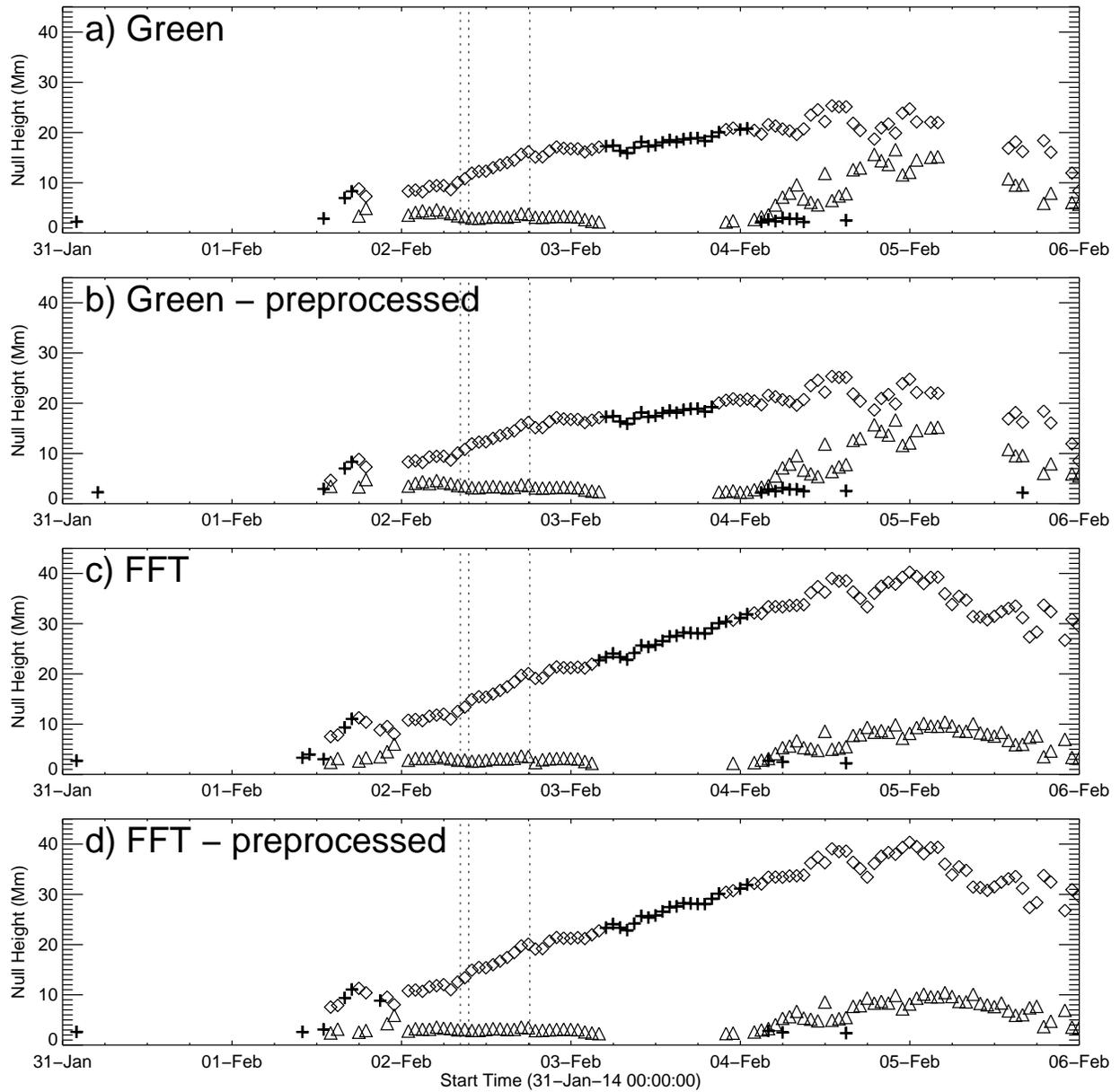}
	\caption{\small Null heights with time. The nulls are identified in potential fields constructed with different methods (Green function vs. fast Fourier transform) and different boundary conditions (original vs. pre-processsed). Vertical dotted lines indicate the peak times of three XMFs. Same as in Figure~\ref{fig:ltc}, the lower (upper) null is marked with a triangle (diamond) for double nulls, while a single null is marked by a `+' symbol.\label{suppfig:nullz}}
\end{figure}

\begin{figure}[!htbp]
	\centering
    \includegraphics[width=\hsize]{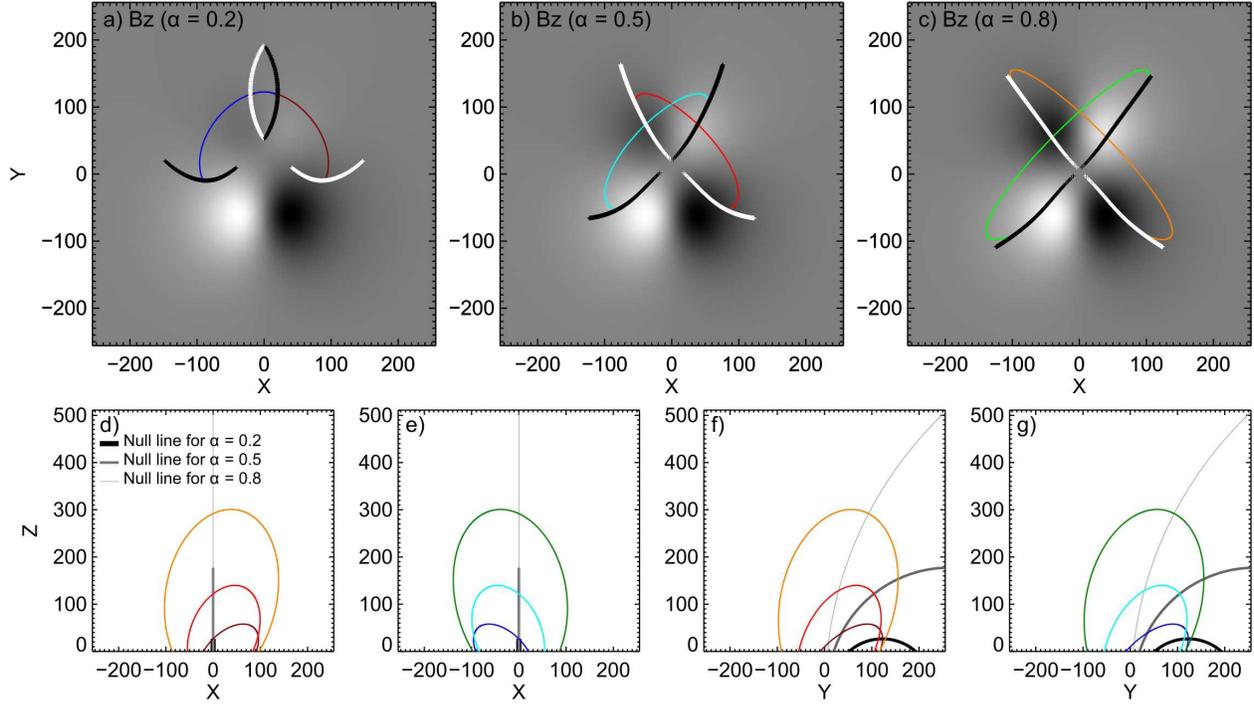}
	\caption{\small Idealized quadrupole field. Panels (a--c) show the map of $B_z$ at $z=0$ with different $\alpha$. The null lines for $\alpha=0.2,\,0.5,\,0.8$ are shown in black, gray, and light gray colors, respectively, with decreasing thickness, in bottom panels. Exemplary spine field lines threading the middle of each null line are also projected in X-Y (a--c), X--Z (d, e), and Y--Z (f, g) planes. The spines resulting from tracing field lines toward the null are shown in maroon ($\alpha=0.2$), red ($\alpha=0.5$), and orange ($\alpha=0.8$), whose footpoints are shown in white in top panels,  while those resulting from tracing field lines away from the null are shown in blue ($\alpha=0.2$), cyan ($\alpha=0.5$), and green ($\alpha=0.8$), whose footpoints are shown in black. \label{suppfig:bquad}}
\end{figure}

\begin{figure}[!htbp]
	\centering
	\includegraphics[width=0.9\hsize]{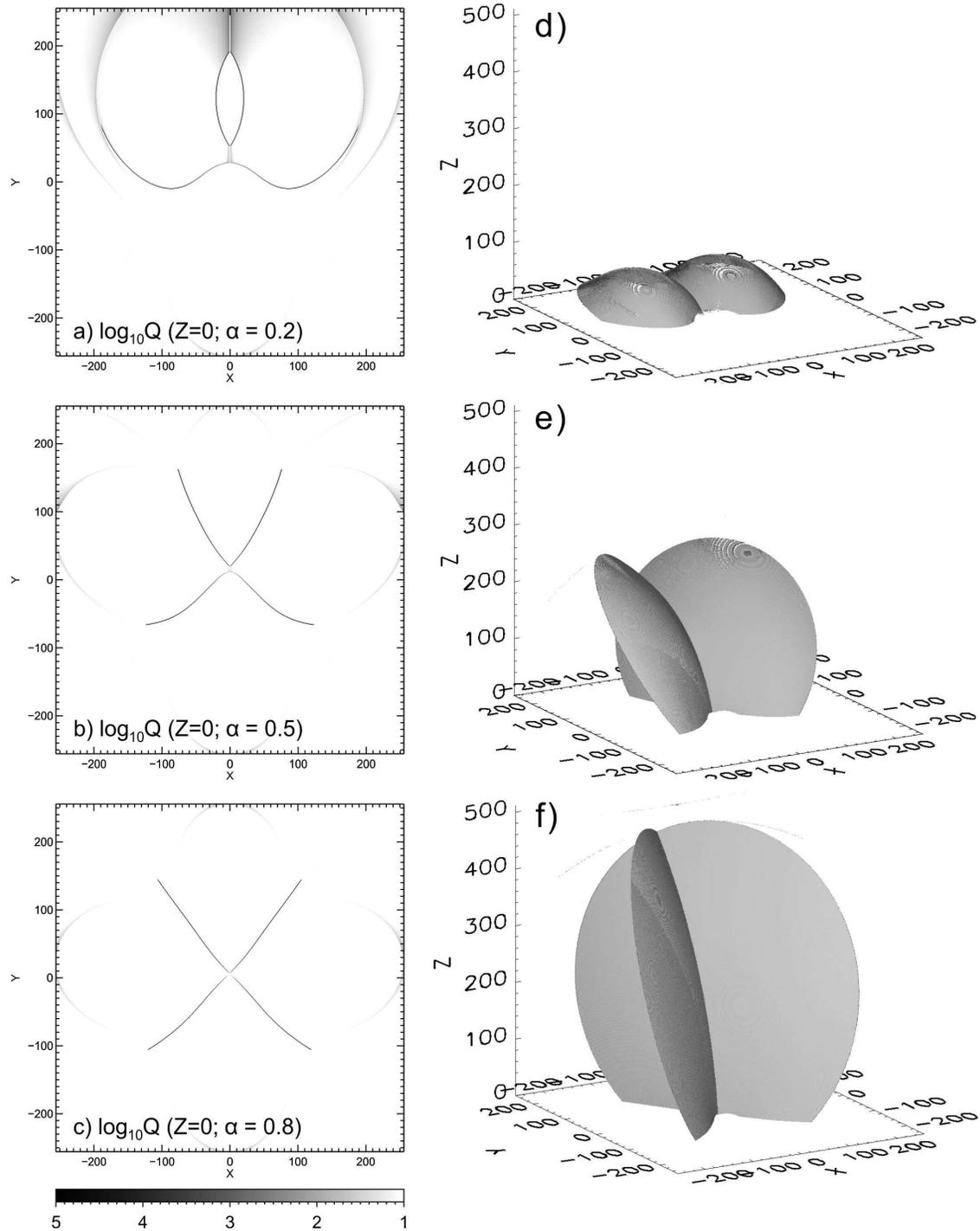}
	\caption{\small Magnetic topology of the quadruple field with different $\alpha$. The left column shows the maps of $\log_{10}Q$ at $z=0$. The corresponding isosurfaces of $\log_{10}Q=4$ in a 3D perspective are shown on the right column. \label{suppfig:toytop}}
\end{figure}

\begin{figure}[!htbp]
	\centering
	\includegraphics[width=\hsize]{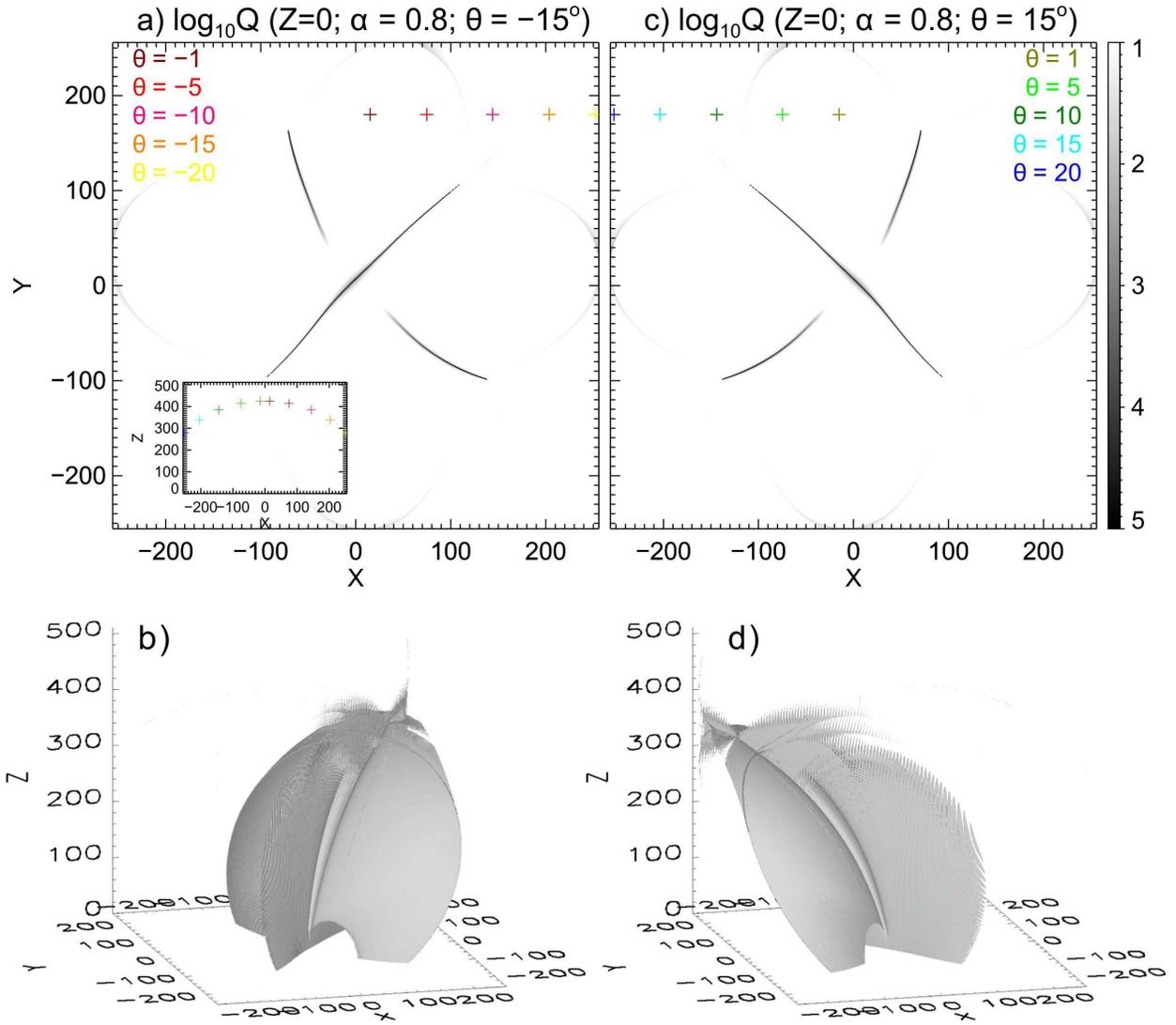}
	\caption{\small Magnetic topology of the quadrupole field when $\mathbf{m}_2 = \alpha\times 1.024\times10^9 \times (\cos\theta,\,\sin\theta,\,0)^T$. Top panels show the maps of $\log_{10}Q$ at $z=0$ when $\theta=-15^\circ$ (a) and $15^\circ$ (c). The corresponding isosurfaces of $\log_{10}Q=4$ are shown in (b) and (d), respectively. The crosses in the top panels mark the null locations with different $\theta$ values (color coded). The inset in (a) shows the null locations in the X-Z plane. \label{suppfig:asym}}
\end{figure}

\begin{figure}[!htbp]
	\centering
	\includegraphics[width=\hsize]{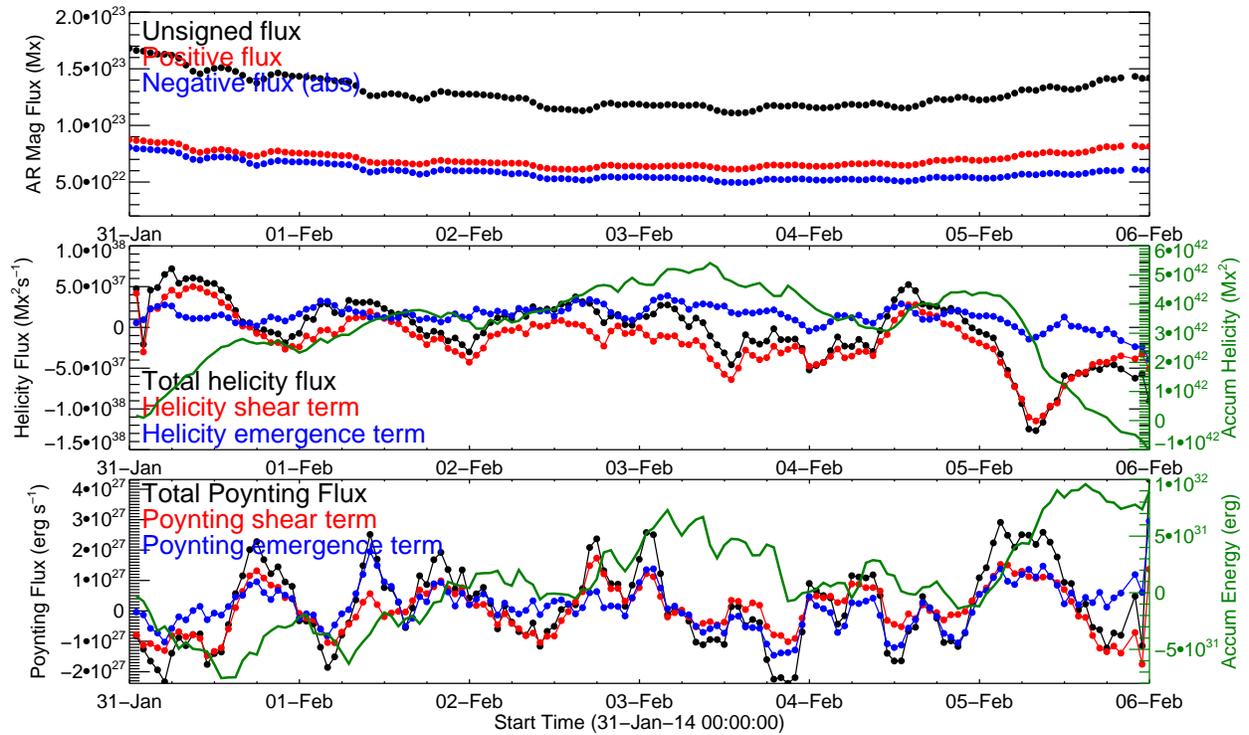}
	\caption{\small Helicity and energy injection into AR 11967. From top to bottom we show the time series of magnetic fluxes [Mx], helicity fluxes [Mx$^2$ s$^{-1}$], and Poynting fluxes [erg s$^{-1}$]. The helicity [Mx$^2$] and magnetic energy [erg] accumulated in the active region are displayed in green in the corresponding panel, scaled by the right y-axis. \label{suppfig:lc}}
\end{figure} 

\clearpage
\section*{Supplementary Movies}
\begin{addendum}
\item [\href{http://staff.ustc.edu.cn/~rliu/preprint/supp_mv1.mov}{Supplementary Movie 1}] AIA observation of the X-shaped major flare (XMF) at 18:11 UT on 2014 February 2 (cf. Figure~\ref{fig:flare3})

\item [\href{http://staff.ustc.edu.cn/~rliu/preprint/supp_mv2.mov}{Supplementary Movie 2}] Evolution of NOAA AR 11967 from January 31 to February 5 (Figure~\ref{fig:bfield})

\item [\href{http://staff.ustc.edu.cn/~rliu/preprint/supp_mv3.mov}{Supplementary Movie 3}] $\log Q$ at different heights for the  potential field at 17:58 UT on 2014 February 2 (cf. Figure~\ref{fig:null}(b--d))

\item [\href{http://staff.ustc.edu.cn/~rliu/preprint/supp_mv4.mov}{Supplementary Movie 4}] 360 deg side view of the isosurfaces of $\log Q =5$ for the potential field at 17:58 UT on 2014 February 2, superimposed by field lines traced in the neighbourhood of two nulls, with fan (spine) field lines in magenta (blue) (cf. Figure~\ref{fig:qsl}).

\item [\href{http://staff.ustc.edu.cn/~rliu/preprint/supp_mv5.mov}{Supplementary Movie 5}] $\log Q$ at different heights for the  NLFFF at 17:58 UT on 2014 February 2 (cf. Figure~\ref{fig:nlff}(c and d))

\item [\href{http://staff.ustc.edu.cn/~rliu/preprint/supp_mv6.mov}{Supplementary Movie 6}] AIA observation of the XMF at 08:20 UT on 2014 February 2 (cf. \suppfig~\ref{suppfig:flare1})

\item [\href{http://staff.ustc.edu.cn/~rliu/preprint/supp_mv7.mov}{Supplementary Movie 7}] AIA observation of the XMF at 09:31 UT on 2014 February 2 (cf. \suppfig~\ref{suppfig:flare2})

\item [\href{http://staff.ustc.edu.cn/~rliu/preprint/supp_mv8.mov}{Supplementary Movie 8}] 360 deg side view of the isosurfaces of $\log Q =5$ for the potential field at 07:58 UT on 2014 February 2, superimposed by field lines traced in the neighbourhood of two nulls, with fan (spine) field lines in magenta (blue) (cf. \suppfig~\ref{suppfig:qsl}(b)).

\item [\href{http://staff.ustc.edu.cn/~rliu/preprint/supp_mv9.mov}{Supplementary Movie 9}] 360 deg side view of the isosurfaces of $\log Q =5$ for the potential field at 08:58 UT on 2014 February 2, superimposed by field lines traced in the neighbourhood of two nulls, with fan (spine) field lines in magenta (blue) (cf. \suppfig~\ref{suppfig:qsl}(d))
\end{addendum}

\end{document}